ARTICLE TYPE

# Galaxy And Mass Assembly: A new approach to quantifying dust in galaxies

B. Farley,[1] U. T. Ahmed,[2,3,4] A. M. Hopkins,[5,3] M. Cowley,[1,6] A. Battisti,[7,8] S. Casura,[9] Y. Gordon,[10] B. W. Holwerda,[11] S. Phillipps,[12] C. Robertson,[13] and T. Zafar[5,3]

[1]School of Chemistry & Physics, Faculty of Science, Queensland University of Technology, Brisbane, QLD 4000, Australia
[2]Australian Astronomical Optics, Macquarie University, 105 Delhi Rd, North Ryde, NSW 2113, Australia
[3]Macquarie University Astrophysics and Space Technologies Research Centre, Sydney, NSW 2109, Australia
[4]Centre for Astrophysics, University of Southern Queensland, 37 Sinnathamby Boulevard, Springfield Central, QLD 4300, Australia
[5]School of Mathematical and Physical Sciences, 12 Wally's Walk Macquarie University, NSW 2109, Australia
[6]University of Southern Queensland, Centre for Astrophysics, West Street, Toowoomba QLD 4350, Australia
[7]ARC Centre of Excellence for All-Sky Astrophysics in 3 Dimensions (ASTRO3D), Australia
[8]Research School of Astronomy and Astrophysics, Australian National University, Canberra, ACT 2611, Australia
[9]Hamburger Sternwarte, Universität Hamburg, Gojenbergsweg 112, 21029 Hamburg, Germany
[10]Department of Physics, University of Wisconsin-Madison, 1150 University Avenue, Madison, WI 53706, USA
[11]Department of Physics and Astronomy, University of Louisville, Natural Science Building 102, Louisville KY 40292, USA
[12]School of Physics, University of Bristol, Tyndall Ave., Bristol BS8 1TL, UK
[13]Department of Physics and Astronomy, University of Louisville, Louisville KY 40292, USA
**Author for correspondence:** B. Farley, Email: breanna.farley@hdr.qut.edu.au.

## Abstract

We introduce a new approach to quantifying dust in galaxies by combining information from the Balmer decrement (BD) and the dust mass ($M_d$). While there is no explicit correlation between these two properties, they jointly probe different aspects of the dust present in galaxies. We explore two new parameters that link BD with $M_d$ by using star formation rate sensitive luminosities at several wavelengths (ultraviolet, H$\alpha$, and far-infrared). This analysis shows that combining the BD and $M_d$ in these ways provides new metrics that are sensitive to the degree of optically thick dust affecting the short wavelength emission. We show how these new "dust geometry" parameters vary as a function of galaxy mass, star formation rate, and specific star formation rate. We demonstrate that they are sensitive probes of the dust geometry in galaxies, and that they support the "maximal foreground screen" model for dust in starburst galaxies.

## 1. Introduction

Fundamental galaxy properties related to star formation, such as the star formation rate (SFR), can be estimated using measurements of H$\alpha$ and ultraviolet (UV) emission (Kennicutt 1998). Both H$\alpha$ and UV wavelengths trace ionising radiation primarily from high-mass OB stars in star-forming regions and are typically more embedded in dust than most of the stellar population. These emissions are susceptible to obscuration caused by dust (Calzetti et al. 2000; Calzetti 2001a, 2001b), which reduces the measured emission at these wavelengths and can lead to underestimation of these properties. Obscuration can be corrected for using methods that employ dust sensitive measurements such as the Balmer Decrement (BD; e.g., Groves, Brinchmann, and Walcher 2012), the UV spectral slope ($\beta$; e.g., Meurer, Heckman, and Calzetti 1999), and the total dust mass ($M_d$) as estimated, for example, by population synthesis tools such as MAGPHYS (Cunha, Charlot, and Elbaz 2008).

Numerous studies have examined the effect of obscuration corrections on the estimates of fundamental galaxy properties. Some explore the properties of a single obscuration correction technique, while others compare and investigate the relationships between various correction methods. For example, Calzetti (2001b) showed how $\beta$ could be used to predict the obscuration of a galaxy and provided an expression to calculate the far-infrared (FIR) to UV ratio as a function of $\beta$. Wijesinghe et al. (2011) studied and compared the BD, $\beta$, and FIR to far-UV (FUV) ratio, and concluded that $\beta$ is a less reliable obscuration indicator due to its sensitivity to other properties. Prior studies also noted limitations with the use of the UV spectral slope. In particular, Kong et al. (2004) and Buat et al. (2005) noted that it was not as reliable a tracer of dust attenuation for galaxies that were not experiencing starbursts. Wang et al. (2016) compared obscuration corrected SFR estimates by using the BD to correct the H$\alpha$ luminosity, $\beta$ to correct the UV luminosity and the UV plus infrared (IR) emission with no correction. This study also concluded that $\beta$ was unlikely to be a reliable obscuration indicator on its own.

The information inferred from an estimate of dust mass, $M_d$, is physically different to the information contained within the BD and $\beta$ as they probe different aspects of the dust properties of a galaxy. BD and $\beta$ are measured using emission sensitive to optical depth, primarily probing optically thin regions. $M_d$ represents the total dust content, including both optically thin and thick regions, in addition to the dust behind the stars which would not be identified as either. However, understanding the distribution of this dust mass (e.g., diffuse interstellar dust compared to clumpy star-forming regions) is essential for accurately judging attenuation. Therefore, while BD and β provide insights into clumpy, star-forming regions, $M_d$ offers a broader view of the dust content, which is crucial for a comprehensive analysis of dust properties in galaxies.

Calzetti, Kinney, and Storchi-Bergmann (1994) studied a sample of starburst or highly star-forming galaxies by applying five different dust geometry models, using the Milky Way (MW) and Large Magellanic Cloud dust extinction laws. Of these cases, only the foreground screen geometry, in which the dust lies in a screen between the observer and the galaxy, was found to be consistent with the data when used with an altered MW extinction law. This analysis made use of the BD as a method of tracing the dust obscuration. They noted that if measurements for the actual dust content of such galaxies were available then it would help in forming a clearer understanding of the impact of dust geometry and chemical composition on the extinction law of these galaxies.

The impact of galaxy inclination on measurements of dust obscuration is well established (Pierini et al. 2004; S. P. Driver et al. 2007). A more inclined (or edge-on) galaxy will have more obscuration due to the increased column of dust along the observer's line of sight. This obscuration, however, depends on the relative star-dust geometry, which is uncertain. Due to geometric effects, different galaxy components, such as bulges and discs (which typically contain different stellar populations), are affected differently, further complicating the problem. The two-component dust model, first suggested by Charlot and Fall (2000) and used in many subsequent studies (Tuffs et al. 2004; Popescu et al. 2011), addresses these complexities by considering both diffuse and clumpy dust components. Recent studies by Lu et al. (2022) and Qin et al. (2024) continue to develop these models. The Chocolate Chip Cookie (CCC) model of Lu et al. (2022) distributes the nebular regions throughout the more diffuse interstellar medium (ISM) like chocolate chips in a cookie. This model successfully describes the effect of inclination on the reddening of both regions, and on the attenuation of Hα, which they measure using the BD. One limitation of the CCC model is that it does not take into account the optically thick star-forming regions, as the BD is only sensitive to the optically thin dust. The model of Qin et al. (2024) is similar to the CCC model in that its two components are the more dense stellar birth clouds and the more diffuse ISM. Qin et al. (2024) used the infrared-to-UV luminosity ratio, referred to as the infrared excess (IRX), to trace the obscuration in their sample of SFGs. Their model is a good fit for their observational data and successfully reproduces the IRX relations. Such inclination effects as reflected in these models are not the focus of this paper, but they are important to acknowledge in order to distinguish and separate them from the "dust geometry" term we use throughout in our analysis.

Popesso et al. (2020) quantified relationships between the BD, metallicity, and inclination angle to serve as proxies for the dust mass and the molecular gas mass, for star-forming galaxies (SFGs) on the main sequence. This was motivated by the need to estimate $M_d$ and molecular gas mass in order to explore their distribution along and across the main sequence of SFGs. Such approaches focus on estimating otherwise unmeasured galaxy properties based on available observables.

Different obscuration indicators, though, such as BD and $M_d$ as used here, are not typically used in combination with one another to infer new information about the dust properties of galaxies. In this analysis we link the BD and $M_d$ together. In essence, this approach uses the BD as a tracer of the optically thin dust, and the $M_d$ as a tracer of the total dust content. Using both jointly enables a deeper understanding of the geometry of the dust in a galaxy. Below, we present new parameters that enable exploration of different aspects of galaxy dust properties, through the joint use of the BD and $M_d$. SFRs and luminosities at FUV, Hα and FIR wavelengths are used to analyse these new parameters.

In § 2 we provide an overview of core concepts related to these new parameters and describe the data used. The new parameters themselves are presented in § 3. In § 4 we present the results and analysis of the new parameters as tracers of dust geometry and optical depth. § 5 discusses these results in relation to fundamental galaxy properties such as stellar mass, SFR and sSFR. Finally, § 6 summarises our findings. Throughout we assume a cosmology with $\Omega_M = 0.3$, $\Omega_\Lambda = 0.7$, and $H_0 = 70$ km s$^{-1}$ Mpc$^{-1}$.

## 2. Quantifying the geometry of dust in galaxies

Shorter wavelengths are more greatly impacted by obscuration due to the small characteristic sizes of dust grains. The degree of obscuration a particular wavelength experiences in a given dust cloud is known as the optical depth of the dust. The optical depth can be characterised by the attenuation parameter, $\tau(\lambda)$, for a simple uniform layer of dust that lies between a source and the observer. This is defined through

$$I_\lambda = I_\lambda^0 e^{-\tau(\lambda)} \tag{1}$$

where $I_\lambda^0$ is the intrinsic intensity of the source, and $I_\lambda$ is the intensity observed (Calzetti, Kinney, and Storchi-Bergmann 1994). The same dust cloud will have a greater optical depth for shorter wavelengths.

The difference in optical depth for the Hα and Hβ emission lines, referred to as the Balmer optical depth, is given by

$$\tau_B^l = \tau_\beta - \tau_\alpha = \ln(\frac{H\alpha/H\beta}{2.86}) \tag{2}$$

(Calzetti, Kinney, and Storchi-Bergmann 1994). The equation from Calzetti, Kinney, and Storchi-Bergmann (1994) used the value 2.88. Here we use 2.86 to remain consistent throughout the paper, although all results change only negligibly if the value 2.88 is used. Equation 2 can be rewritten as

$$\tau_B^l = 2.303 \log(\frac{H\alpha}{H\beta}) - 1.0508. \tag{3}$$

The geometry of dust with respect to stars is known to have an impact on the observed level of obscuration affecting a galaxy's emission (e.g., Tuffs et al. 2004; Natale et al. 2015; Narayanan et al. 2018; Lin et al. 2021; Sachdeva and Nath 2022; Witt, Thronson, and Capuano 1992). In Figure 1, we present illustrations depicting two extreme versions of a dust geometry model. The first such model considered here, known as the foreground screen dust geometry model (Calzetti, Kinney, and Storchi-Bergmann 1994), assumes that all the dust lies in a screen between the observer and the stars (Figure 1a–c). Figure 1a shows the case in which there is a low $M_d$ in a foreground screen geometry. In this scenario, there is little obscuration of the light. In Figure 1b, the increase in $M_d$ results in greater obscuration due to the increased optical depth. In Figure 1c the $M_d$ has increased to an extreme limit in which the optical depth is so great that any starlight is completely obscured.

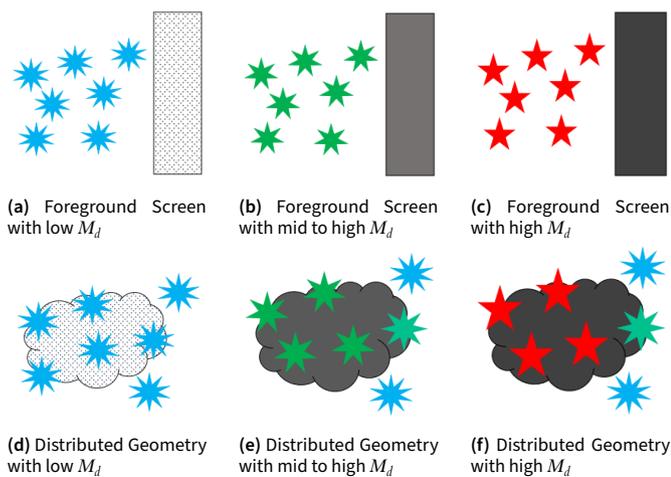

**Figure 1.** Diagram of different dust geometries. The observer lies to the right of each panel in this figure. Lightly coloured dust screens or clouds indicate low $M_d$, and darker dust screens or clouds indicate higher $M_d$. Blue stars (with many points) indicate little or no obscuration, green stars indicate mid to high obscuration, and red stars (with five points) indicate complete obscuration. Each panel in this figure represents a galaxy of the same size, such that an increase in dust mass results in an increase in optical depth.

In the distributed dust geometry model the dust and stars are mixed (Figure 1d–f). Figure 1d illustrates a scenario with low $M_d$ in a distributed geometry, resulting in little obscuration of the light. In Figure 1e the $M_d$, and therefore optical depth, are increased. The light from deeper within the cloud is more obscured due to travelling through more dust to reach the observer. Conversely, the light from the edges of the cloud closer to the observer are less obscured. In Figure 1f the $M_d$ has been increased to an observational limit in which the light from deeper within the cloud is no longer detected. This light has been completely obscured. The light from the closer edges of the cloud, which travels through less dust, is less obscured, resulting in the detection of some obscured light.

With the foreground screen geometry, all light of a given wavelength experiences consistent levels of obscuration, whether little or heavy obscuration, as it must travel through the same amount of dust. With a distributed dust geometry, light of a given wavelength from stars deeper within the cloud will experience more obscuration than light of the same wavelength from stars towards the edges of the cloud as the light from deeper within the cloud must travel through more dust. This means that the relative depth of stars within the dust cloud determines the level of obscuration, with deeper stars experiencing higher attenuation. Additionally, as the $M_d$ increases, a wider range of obscuration levels can be detected due to the varying positions of stars within the dust clouds.

The level of obscuration within galaxies can be quantitatively assessed through the BD, while the optical depth is intrinsically connected to $M_d$. Given this relationship, combining the BD and $M_d$ offers a promising approach to investigate the intricate geometry of dust in galaxies. We use data from the Galaxy And Mass Assembly (GAMA) survey, described below, following which we introduce two novel parameters that link the BD and $M_d$. These parameters aim to provide deeper insights into the role of dust geometry in influencing dust properties and star formation rates.

## 2.1 Data

The GAMA survey is a spectroscopic and photometric survey that covers approximately 250 deg$^2$ of the sky over 5 regions using the AAOmega spectrograph of the 3.9m Anglo-Australian Telescope (S. P. Driver et al. 2011; Liske et al. 2015; Simon P. Driver et al. 2016; Baldry et al. 2018; Simon P Driver et al. 2022).

The GAMA data and derived parameters are organised into Data Management Units (DMUs, Liske et al. 2015). We use data and derived parameters from various DMUs, detailed in Table 1. Specifically, emission line flux measurements and uncertainties for H$\alpha$, H$\beta$, N[II], and O[III], in addition to equivalent widths measurements and uncertainties for the H$\alpha$ and H$\beta$ emission lines are obtained from the SpecLineSFR DMU (Gordon et al. 2017). The SpecLineSFR DMU also provides the original survey source and redshift estimates with redshift quality measures, nQ. The MAGPHYS DMU provides dust mass, $M_d$, estimates and percentile ranges which were used to obtain the uncertainties on $M_d$. While MAGPHYS is a robust tool for estimating dust masses, it is important to acknowledge that different SED models can yield systematically different $M_d$ estimates. The StellarMasses DMU (Taylor, Hopkins, Baldry, Brown, Driver, Kelvin, Hill, Robotham, Bland-Hawthorn, Jones, Sharp, Thomas, Liske, Loveday, Norberg, Peacock, Bamford, Brough, Colless, Cameron, Conselice, Croom, Frenk, Gunawardhana, Kuijken, Nichol, Parkinson, Phillipps, Pimbblet, Popescu, Prescott, Sutherland, Tuffs, van Kampen, et al. 2011b) provides the stellar mass, $M_*$, estimates and uncertainties in addition to obscuration-corrected absolute magnitude measurements and uncertainties in the $r$ band. The $M_*$ estimates of Taylor, Hopkins, Baldry, Brown, Driver, Kelvin, Hill, Robotham, Bland-Hawthorn, Jones, Sharp, Thomas, Liske, Loveday, Norberg, Peacock, Bamford, Brough, Colless, Cameron, Conselice, Croom, Frenk, Gunawardhana, Kuijken, Nichol, Parkinson, Phillipps, Pimbblet, Popescu, Prescott, Sutherland, Tuffs, van Kampen, et al. (2011b) are based on the (Bruzual and Charlot 2003) population synthesis code, which is also used by MAGPHYS (Cunha, Charlot, and El-baz 2008; Taylor, Hopkins, Baldry, Brown, Driver, Kelvin, Hill, Robotham, Bland–Hawthorn, Jones, Sharp, Thomas, Liske, Loveday, Norberg, Peacock, Bamford, Brough, Colless, Cameron, Conselice, Croom, Frenk, Gunawardhana, Kuijken, Nichol, Parkinson, Phillipps, Pimbblet, Popescu, Prescott, Sutherland, Tuffs, Kampen, et al. 2011a). Uncorrected measurements and uncertainties for the *r* band and FUV band were also used. The FIR flux measurements and uncertainties were taken from the gkvFarIR DMU (Bellstedt et al. 2020b). The approximate elliptical semi-major axis and axial ratio values used to calculate the galaxy areas were obtained from the gkvInputCat DMU (Bellstedt et al. 2020a).

**Table 1.** Summary of the data and derived parameters used and their GAMA DMUs.

| DMU | Data & Derived Parameters |
| --- | --- |
| SpecLineSFR v05 | Z, NQ, SURVEY, SURVEY_CODE, HA_FLUX, HA_FLUX_ERR, HA_EW, HA_EW_ERR, HB_FLUX, HB_FLUX_ERR, HB_EW, HB_EW_ERR, NIIR_FLUX, NIIR_FLUX_ERR, OIIIR_FLUX, OIIIR_FLUX_ERR, OIIIB_FLUX, OIIIB_FLUX_ERR |
| MAGPHYS v06 | mass_dust_best_fit, mass_dust_percentile16, mass_dust_percentile84 |
| StellarMasses v24 | mstar, delmstar, absmag_FUV, delabsmag_FUV, absmag_r, delabsmag_r, absmag_r_stars, delabsmag_r_stars |
| gkvFarIR v03 | FIR_flux_PSF_p100, FIR_flux_PSF_err_p100 |
| gkvInputCat v02 | R50, axrat |

The sample used here was selected to ensure the data was of high quality. Only objects with redshifts originating from the GAMA, SDSS and 2dFGRS surveys were used. The sample was limited to objects with redshift quality nQ ≥ 3, which indicates a reliable estimate that is suitable for use in scientific analysis (S. P. Driver et al. 2011). Only galaxies with lines in emission were retained, and the emission line flux measurements (Hα, Hβ, N[II], O[III]) were constrained to only those with a signal-to-noise ratio (S/N) ≥ 5. Similarly, the sample only contains dust mass estimates with S/N ≥ 3, and FIR flux measurements with S/N ≥ 1. We use the standard optical diagnostic diagram of Baldwin, Phillips, and Terlevich (1981) (BPT) to classify our galaxies as star forming or hosting active galactic nuclei (AGN). AGN were identified using the Kauffmann et al. (2003) definition, and removed from our sample due to the significant effect AGN can have on the observed emission lines and the inferred properties of dust within galaxies. Table 2 shows how many objects remain after these various criteria are applied. Highly inclined galaxies are not removed from this sample. Less than 5% of the galaxies in this sample are moderately or highly inclined (with an axial ratio < 0.4), and removing these galaxies does not alter our results. Retaining these more highly inclined galaxies maintains the completeness of the sample and demonstrates that our analysis is not sensitive to inclination.

**Table 2.** The number of objects remaining in the sample after each selection criteria was applied to the data.

| Selection Criteria | Number of Objects |
| --- | --- |
| None | 194 053 |
| Survey (GAMA, SDSS & 2dFGRS) | 192 213 |
| Redshift Quality (nQ) | 182 981 |
| Emission Lines & FIR S/N | 3 756 |
| Dust Mass S/N | 1 141 |
| SFGs via BPT Diagram | 842 |

## 2.2   BD, $M_d$ & Dust Geometry

To quantify the BD we use the Hα and Hβ emission line flux measurements. These were first corrected for stellar absorption following Hopkins et al. (2001), as

$$F_{H\alpha} = \frac{(H\alpha EW + EW_c)}{H\alpha EW} f_{H\alpha} \qquad (4)$$

$$F_{H\beta} = \frac{(H\beta EW + EW_c)}{H\beta EW} f_{H\beta} \qquad (5)$$

where $f_{H\alpha}$ is the Hα emission line flux, $f_{H\beta}$ is the Hβ emission line flux, HαEW is the equivalent width of the Hα emission line, HβEW is the equivalent width of the Hβ emission line, and we adopt the same stellar absorption equivalent width correction for both Hα and Hβ of $EW_c = 2.5$ following Gunawardhana et al. (2013). Finally, the BD values are calculated as

$$BD = \frac{F_{H\alpha}}{F_{H\beta}}. \qquad (6)$$

The BD values calculated from each of GAMA, SDSS & 2dFGRS were compared to ensure that all three were providing consistent BD values within acceptable ranges. Although 2dF-GRS spectra are not flux calibrated, the BD values are reliable, as they span the same range as those seen with the flux calibrated GAMA and SDSS spectra. If we omit them from our analysis our results remain unchanged apart from having slightly fewer galaxies represented.

Figure 2 shows the relationship between the BD and both $M_d$ (Figure 2a) and dust surface density ($\Sigma_{M_d}$) (Figure 2b). There is an upper envelope for the observed BD values evident, which increases with increasing $M_d$ and $\Sigma_{M_d}$. The dotted lines shown tracing these envelopes are empirical characterisations. For this dataset, the two different envelope lines are given by:

$$\log(BD_{Env}) = 0.185 \log(M_d) - 0.481, \qquad (7)$$

and

$$\log(BD_{Env}) = 0.459 \log(\Sigma_{M_d}) - 1.801. \qquad (8)$$

It is important to note that if different samples are being used, while qualitatively similar, such envelopes may differ quantitatively, especially if $M_d$ is estimated with different population synthesis tools.

The envelope is not a consequence of the S/N limits placed on the dataset, as an envelope of the same shape is present even when no S/N limits are imposed. Investigating the Hβ line flux as a function of $M_d$ does reveal that at the lowest values of $M_d$ there is a tendency for the faintest Hβ fluxes to be absent. This, however, does not explain the existence of the envelope at the high $M_d$ end, nor its relatively linear shape over the full range of $M_d$. As a result, we are confident that the envelope seen here is not an observational bias, nor is it a result of our sample selection limits.

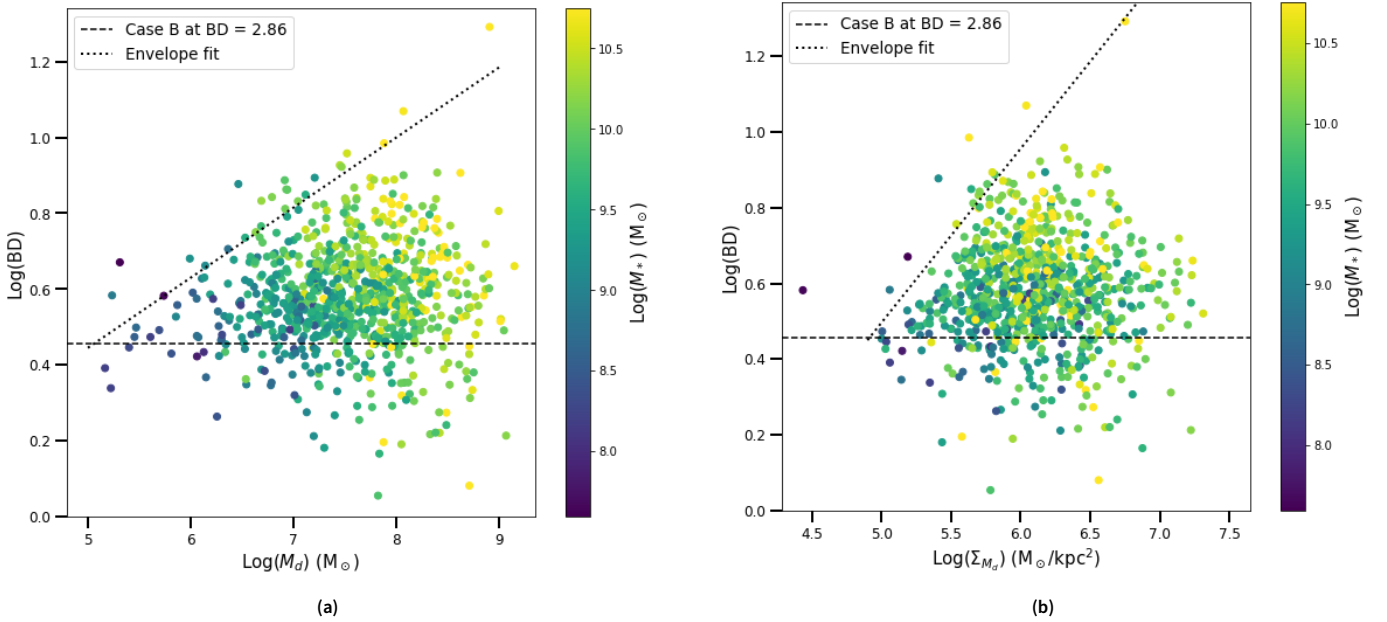

(a)　　　　　　　　　(b)

**Figure 2.** BD as a function of (a) $M_d$, and (b) dust surface density, coloured by $M_*$. The black dotted line represents the observed upper envelope of the data. The black dashed line represents the BD Case B value of 2.86. This Case B value of 2.86 is the BD value which corresponds to no obscuration (Osterbrock 1989). The correlation coefficient for panel (a) is 0.022 and the correlation coefficient for panel (b) is 0.013.

In the case of Figure 2a this envelope traces the BD values resulting from a foreground screen geometry as the optical depth of the screen increases. We focus here on BD and $M_d$ for the purpose of this illustrative discussion of dust geometry. The low optical depth foreground screen geometry lies at the low BD, low $M_d$ end of the envelope. The high optical depth foreground screen geometry lies at the high BD, high $M_d$ end of the envelope. The models from Figure 1 are positioned in Figure 3, to capture their representative locations in the diagram, and to emphasise how each model would be reflected in the quantitative measurements.

In the case of low dust content, the foreground screen geometry and distributed geometry both fall in the low BD region, as neither provides enough attenuation to produce high BD measurements. This indicates that with less dust, the differences in geometries become more redundant in terms of their impact on the BD. However, even in low dust content scenarios, certain configurations of distributed geometry, such as those depicted in Figure 1f, can result in variations in the BD values.

There is a point at which the dust becomes so optically thick that the Balmer lines are too attenuated to escape. In the case of the foreground screen this results in no measurement of the Balmer lines at all, and corresponds to the high BD,

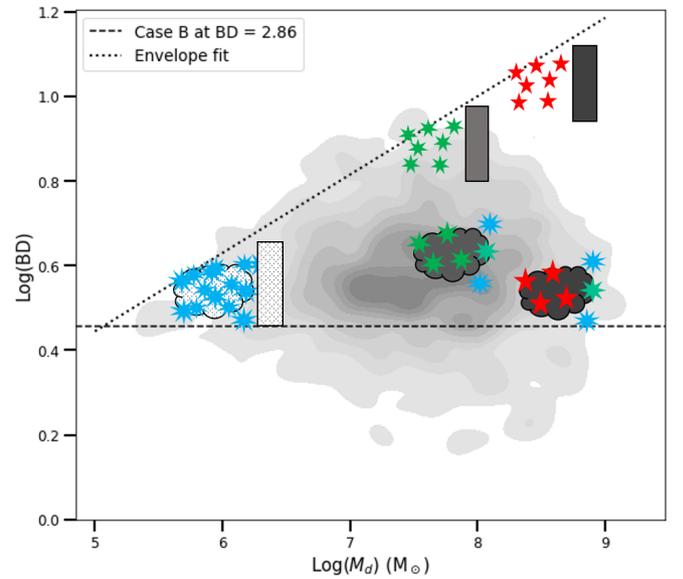

**Figure 3.** Schematic diagram conceptualising where the different dust geometries shown in Figure 1 are expected to fall in the BD vs $M_d$ diagram. The data from Figure 2a is represented here as a grey data density plot.

high $M_d$ region that is lacking any data, as seen in the upper right of Figures 2a and 3.

For the case of the distributed dust geometry, the emission lines from stars buried within the dust cloud escape less than emission at the outer boundary. This results in the emission lines originating in the edges of the cloud experiencing less attenuation and being observed, allowing for measurements that lie in the low BD, high $M_d$ region. Therefore, the data points along the Case B line, with a low BD, correspond to a maximal distributed dust geometry. A mixture of these two extreme dust geometries allows for the spread of values observed between the envelope line and the Case B line.

It is well established that $M_d$ in galaxies is correlated with the $M_*$. This is less the case for $\Sigma_{M_d}$, and these relationships are shown together in Figure 4. Panel 4a shows the direct relationship between stellar mass and $M_d$, while panel 4c shows the ratio of $M_d$ to stellar mass as a function of stellar mass. Panel 4b displays the relationship between $\Sigma_{M_d}$ and stellar mass. There is no panel that depicts the relationship between $\Sigma_{M_d}$ and $\Sigma_{M_*}$ as that would in effect be the same as panel 4a. However, panel 4d does show the ratio of $\Sigma_{M_d}$ to $\Sigma_{M_*}$ as a function of $\Sigma_{M_*}$. These relationships are explored further by Cortese et al. (2012), Clemens et al. (2013), Calura et al. (2017), De Vis et al. (2017), Orellana et al. (2017), V. Casasola et al. (2020), and Viviana Casasola et al. (2022). The trends in Figure 4 are colour coded by the SFR estimated from the Hα luminosity, which we calculate as follows. The obscuration corrected $L_{H\alpha}$ comes from

$$L_{H\alpha} = (EW_{H\alpha} + EW_c)10^{-0.4(M_r - 34.10)}$$
$$\times \frac{3 \times 10^{18}}{[6564.61(1+z)]^2}\left(\frac{BD}{2.86}\right)^{2.36} \quad (9)$$

where $M_r$ is the obscuration corrected $r$ band absolute magnitude, and $z$ is the redshift (Gunawardhana et al. 2011; Gunawardhana et al. 2013). If the observed BD is less than 2.86, then it is set to 2.86 here, equivalent to having no obscuration correction term. Equation 9 also includes terms that apply aperture and stellar absorption corrections, which are required for the Hα. The $SFR_{H\alpha}$ is then calculated as

$$SFR_{H\alpha} = \frac{L_{H\alpha}}{1.27 \times 10^{34}} \quad (10)$$

(Kennicutt 1998; Gunawardhana et al. 2011).

The $M_d$ and stellar mass are correlated, although with a scatter of about 1 dex around the broad trend. This is reflected in the relatively flat relationship seen in Figure 4c, albeit with the scatter emphasised in this representation. It is apparent that much of this scatter is related to the SFR. As the SFR increases with stellar mass, galaxies of a given dust mass with high SFRs have larger stellar masses than those with low SFRs. Figure 4d shows the ratio of surface densities for dust and stellar mass, which is actually identical to $M_d/M_*$ as the surface area term cancels. Looking at this parameter as a function of $\Sigma_{M_*}$, however, highlights that galaxies with the largest stellar mass surface density favour a lower proportion of dust mass, although the trend is mostly driven by the relatively small number of galaxies with $\Sigma_{M_*} \gtrsim 10^9 M_\odot/\text{kpc}^2$. This may suggest that the most compact galaxies may have proportionally less dust than more typical star forming galaxies. The interplay between stellar mass, $M_d$, and SFR implicitly includes the contribution of BD, as it is incorporated in the Hα SFR estimate. In order to tease out these related parameters further, it is helpful to explore new ways of quantifying the links between BD, $M_d$, and $\Sigma_{M_d}$. We start with an investigation of SFR estimators that are sensitive in different degrees to the presence of obscuring dust.

## 3. The role of BD in understanding SFR tracers

Koyama et al. (2015) studied the relationship between Hα attenuation and the ratio of $SFR_{H\alpha}$ to $SFR_{FUV}$. This ratio can highlight the effect of optical depth due to the different effect at the different wavelengths. When the dust is more optically thick, proportionally less of the FUV emission will be detected, resulting in higher values of $SFR_{H\alpha}/SFR_{FUV}$. Koyama et al. (2015) found a positive correlation between the Hα attenuation and $SFR_{H\alpha}/SFR_{FUV}$, but noted that there is substantial scatter surrounding this relationship. Due to the scatter, they determined that dust attenuation levels could be roughly estimated using $SFR_{H\alpha}/SFR_{FUV}$, but that it could not be used as a more precise method of deriving the dust attenuation.

We reproduced their figure comparing Hα attenuation and $SFR_{H\alpha}/SFR_{FUV}$, but instead used the BD rather than the Hα attenuation, as a starting point to explore the relationships surrounding dust geometry and optical depth(Figure 5a). This figure shows the relationship when no obscuration corrections are applied to the SFR estimates. For $SFR_{H\alpha,Obs}$, this simply corresponds to omitting the BD term from Equation 9. For $SFR_{FUV,Obs}$ we use

$$L_{FUV,Obs} = 4\pi D_L^2 \times 10^{-0.4(m_{AB,FUV} + 56.1)} \quad (11)$$

where $D_L$ is the luminosity distance, $m_{AB,FUV}$ is the FUV band apparent AB magnitude. This is converted to SFR through

$$SFR_{FUV,Obs} = \frac{L_{FUV,Obs}}{7.14 \times 10^{20}} \quad (12)$$

(Kennicutt 1998; Hopkins et al. 2003).

Figure 5a shows a positive correlation between the BD and $SFR_{H\alpha,Obs}/SFR_{FUV,Obs}$. This is expected as both are tracers of the obscuration present, consistent with the results of Koyama et al. (2015). To implement an obscuration correction for $SFR_{FUV}$, we follow Calzetti et al. (2000), using

$$L_{FUV}(\lambda) = L_{FUV,Obs}(\lambda)10^{0.4E(B-V)k(\lambda)} \quad (13)$$

(Hopkins et al. 2001) where $k(\lambda)$ is the reddening curve (Calzetti et al. 2000). For FUV the wavelength is $\lambda = 1500$ Å $= 0.15$ μm and $k(FUV) = 10.33$. $E(B-V)$ is given by

$$E(B-V) = 0.44\frac{\log(BD)}{0.4(k(H\beta) - k(H\alpha))} \quad (14)$$

(Calzetti 2001a; Hopkins et al. 2001), with $k(H\alpha) = 2.38$ and $k(H\beta) = 3.65$. The obscuration corrected $SFR_{FUV}$ values are again calculated with the same SFR calibration factor

$$SFR_{FUV} = \frac{L_{FUV}}{7.14 \times 10^{20}} \quad (15)$$

(Kennicutt 1998; Hopkins et al. 2003).

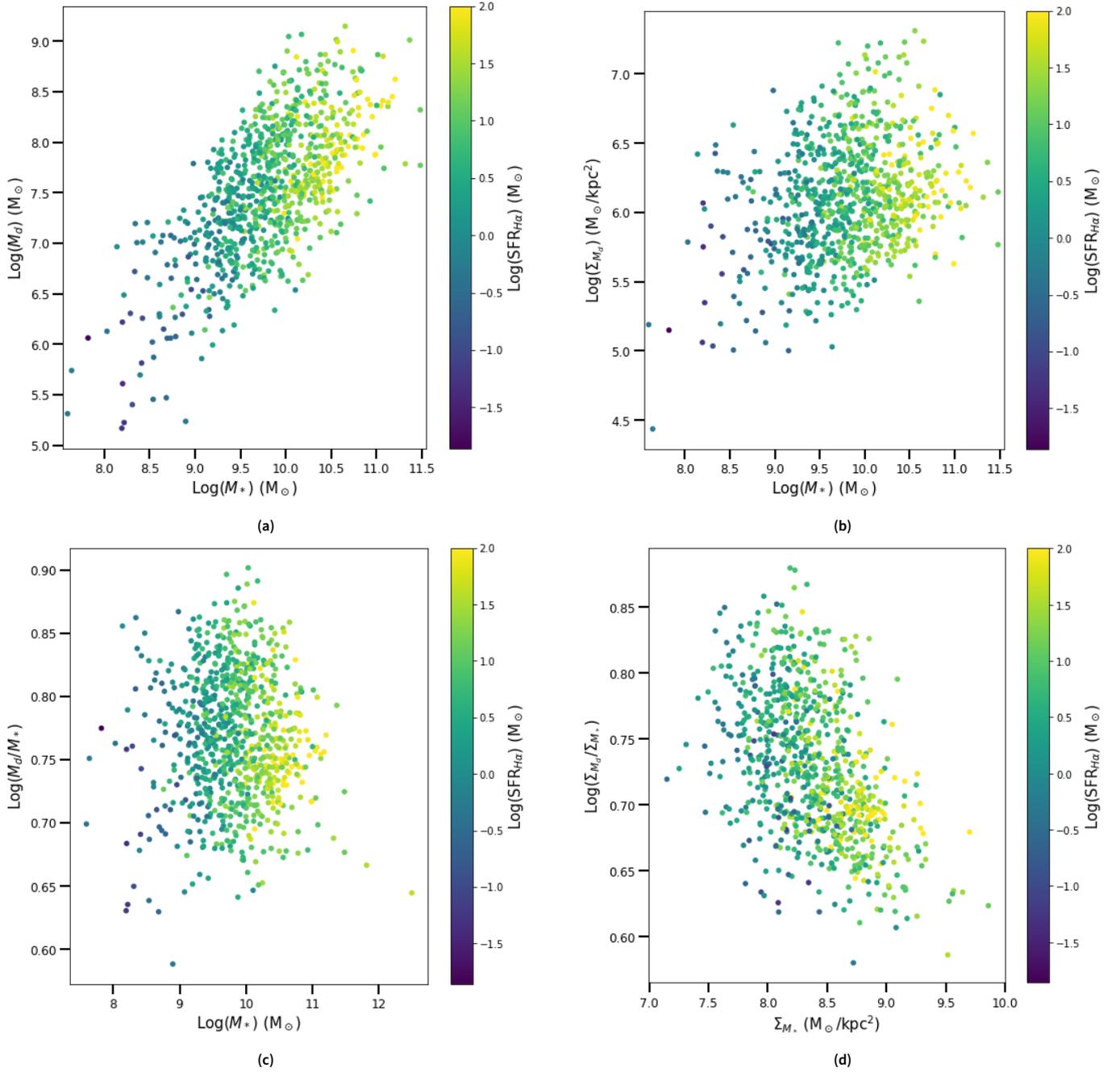

**Figure 4.** (a) $M_d$ as a function of $M_*$, (b) $\Sigma_{M_d}$ as a function of $M_*$, (c) $M_d/M_*$ as a function of $M_*$, and (d) $\Sigma_{M_d}/\Sigma_{M_*}$ as a function of $\Sigma_{M_*}$, with all panels coloured by H$\alpha$ SFR. The correlation coefficients are 0.66 for panel (a), 0.28 for panel (b), -0.032 for panel (c), and -0.47 for panel (d).

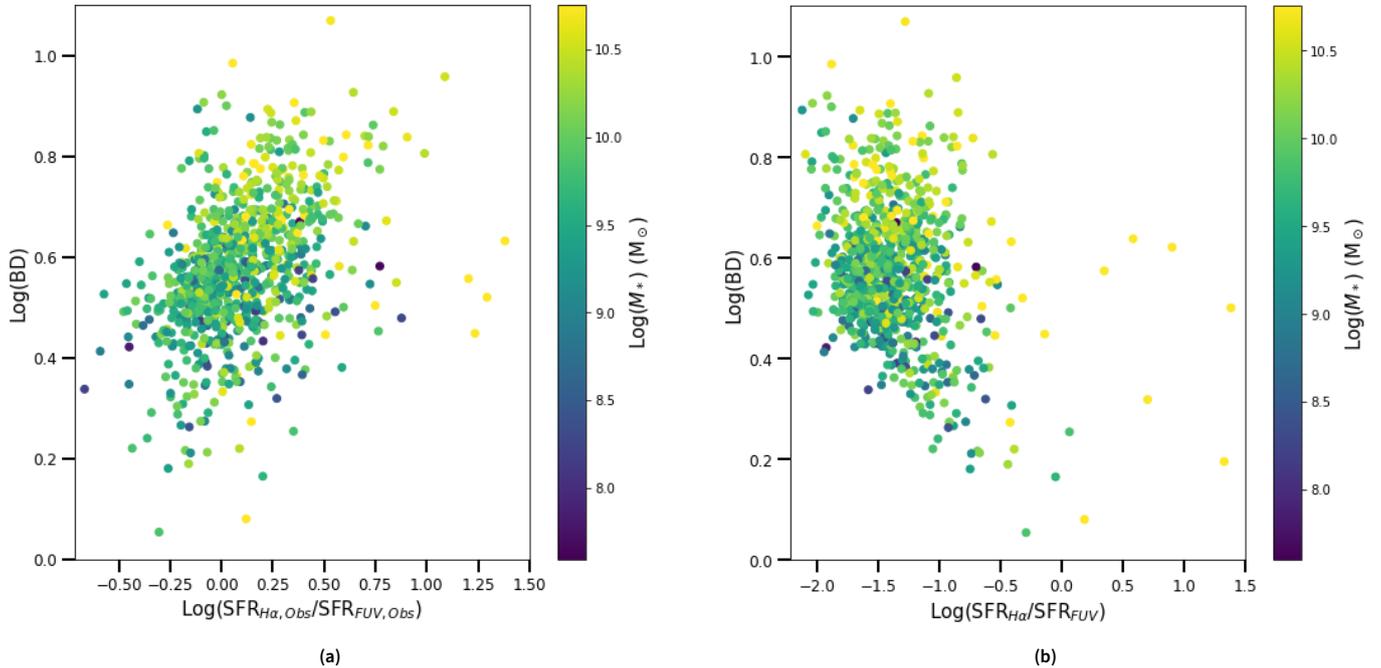

**Figure 5.** (a) BD as a function of $SFR_{H\alpha,Obs}/SFR_{FUV,Obs}$, and (b) BD as a function of $SFR_{H\alpha}/SFR_{FUV}$, both coloured by $M_*$. The correlation coefficient for panel (a) is 0.21 and the correlation coefficient for panel (b) is -0.399.

With obscuration corrections in place, the correlation between BD and $SFR_{H\alpha}/SFR_{FUV}$ is no longer present (Figure 5b). This difference when the obscuration corrections are applied is a direct consequence of the degree of obscuration correction, and that the FUV measurements require a larger correction for a given value of BD compared to the H$\alpha$ measurements. This is reflected in the apparent lower diagonal envelope to the data distribution, where higher values of BD lead to proportionally higher $SFR_{FUV}$ compared to $SFR_{H\alpha}$. This causes the SFR ratio to decrease as BD increases, changing the positive correlation with the uncorrected ratio to a more or less vertical, uncorrelated, distribution. There is also a trend for galaxies of higher $M_*$ to show higher BD, seen in the vertical colour gradient. This is expected from the fact that higher $M_*$ implies higher $M_d$, and high values of BD are only seen in high $M_d$ galaxies.

Instead of just investigating the H$\alpha$ and UV luminosities and their ratios, we can introduce other wavelength measurements as well. In particular the FIR emission is a well known tracer of the total dust-reradiated emission. We now use it as well in exploring this approach. Here we refer to the ratio of the FIR luminosity to the BD corrected H$\alpha$ luminosity as the "H$\alpha$ deficit." We choose this term since, if some degree of optically thick dust is present, the H$\alpha$ luminosity will be reduced in comparison to the FIR luminosity. In the absence of optically thick dust affecting the H$\alpha$, the two should be proportional, both tracing the underlying SFR.

The FIR luminosity is calculated simply with

$$L_{FIR} = 4\pi D_L^2 \times f_{FIR} \qquad (16)$$

where $f_{FIR}$ is the *Herschel*-ATLAS 100 $\mu$m flux, and $D_L$ is the luminosity distance. These results are qualitatively unchanged if we use the 160 $\mu$m flux instead, or a linear combination of both. We choose to present the results here in terms of a single FIR band for simplicity. At this stage, we are ready to return to quantifying links between the BD, $M_d$, and $\Sigma_{M_d}$.

## 4. A New Approach

To link the BD with the $M_d$ and $\Sigma_{M_d}$, we introduce two new parameters. The first of these is a new parameter aimed at quantifying the mixing of the foreground screen and distributed dust geometries. This parameter, $F_{dust}$, is calculated as

$$F_{dust} = \frac{\log BD - \log(2.86)}{\log BD_{Env} - \log(2.86)} \qquad (17)$$

where $BD_{Env}$ is the BD value at the envelope line from Figure 2a (Equation 7).

$F_{dust}$ quantifies where a BD value lies vertically in relation to the Case B and envelope lines. In doing so, it quantifies the proportion that each geometry contributes in that galaxy, with higher values of $F_{dust}$ indicating a more foreground screen geometry and lower values of $F_{dust}$ indicating a more distributed geometry. A value of $F_{dust} = 1$ may be referred to as a "maximal foreground screen" geometry, and $F_{dust} = 0$ as a "maximal distributed dust" geometry. In our sample, for the small number of galaxies with BD lying above the envelope line or below the Case B line, we set $F_{dust}$ to be 1 or 0 respectively.

It is important to acknowledge that the terms "foreground screen" and "distributed dust" are used as convenient descriptors of dust distribution regimes and should not be interpreted literally. The primary distinction is that the "distributed dust" regime can result in significant portions of the SFR and associated Balmer lines being entirely obscured by dust, while other regions experience only slight extinction. As a result, the BD can be small despite substantial loss of Hα emission. In contrast, the "foreground screen" regime implies a moderate and more uniform obscuration of all emission regions, leading to a larger BD while still allowing a considerable amount of Hα emission to be observed.

Since Figure 2b shows a similar structure with an envelope line, a version of $F_{\text{dust}}$ may also be calculated from this $\Sigma_{M_d}$ version of the figure. The version of $F_{\text{dust}}$ calculated from the $\Sigma_{M_d}$ figure (Figure 2b) will be referred to as $\Sigma_{F_{\text{dust}}}$ and uses $\text{BD}_{\text{Env}}$ from Equation 8.

Figure 6 compares the $F_{\text{dust}}$ and $\Sigma_{F_{\text{dust}}}$ values. The data are centred quite evenly around the 1:1 line, although there is a slight tendency towards somewhat higher $F_{\text{dust}}$ values compared to $\Sigma_{F_{\text{dust}}}$. The mostly even distribution about the 1:1 line indicates that $F_{\text{dust}}$ and $\Sigma_{F_{\text{dust}}}$ are sufficiently similar that either may be used in exploring dust geometry. However, it is worth noting that the interpretation of the envelope line in these diagrams cannot imply the same dust geometry in both cases. We argue that $F_{\text{dust}}$ is the choice that better matches a model where the envelope line represents a foreground screen dust geometry.

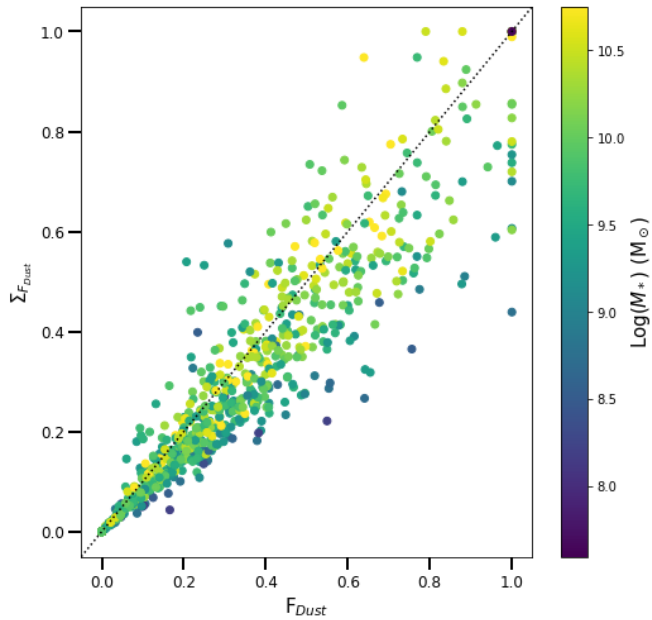

**Figure 6.** $\Sigma_{F_{\text{dust}}}$ as a function of $F_{\text{dust}}$, coloured by $M_*$. The dotted line is a 1:1 line. The correlation coefficient is 0.94.

Independent of any envelope line, the BD and $M_d$ may still be combined in such a way that more information about the geometry and optical depth of the dust may be inferred than if they are used independently. We introduce a second new parameter, $H_{\text{dust}}$, that links the BD and $M_d$ in a different way, in order to quantify the dust geometry. $H_{\text{dust}}$ is defined in terms of the BD as

$$H_{\text{dust}} = 10^{1.0508} \frac{M_d}{\text{BD}^{2.303}}. \tag{18}$$

Equation 18 was derived using

$$H_{\text{dust}} = \frac{M_d}{10^{5} \tau_B^f} \tag{19}$$

which is equivalent in the case of a foreground screen geometry. Therefore, $H_{\text{dust}}$ can be interpreted as a normalised dust mass, providing a way to quantify dust geometry beyond using BD and $M_d$ independently. As with $F_{\text{dust}}$, an alternate version of $H_{\text{dust}}$ can be calculated in which $\Sigma_{M_d}$ is used in place of $M_d$. This dust surface density version will be referred to as $\Sigma_{H_{\text{dust}}}$.

Figure 7 compares $H_{\text{dust}}$ and $\Sigma_{H_{\text{dust}}}$. Although the $H_{\text{dust}}$ values are consistently higher than the $\Sigma_{H_{\text{dust}}}$ values, the two parameters are still correlated. For the purposes of this investigation $H_{\text{dust}}$ is the more intuitive quantity. $H_{\text{dust}}$ is a global parameter as it uses the total dust content of the galaxy, whereas $\Sigma_{H_{\text{dust}}}$, using a surface density, is related to the spatial distribution of the dust. In this analysis we are interested in exploring parameters like star formation and dust geometry as global galaxy properties. Were another study to be conducted focusing on surface densities and the spatial distribution of galaxy properties, then perhaps $\Sigma_{H_{\text{dust}}}$ would be the more useful parameter.

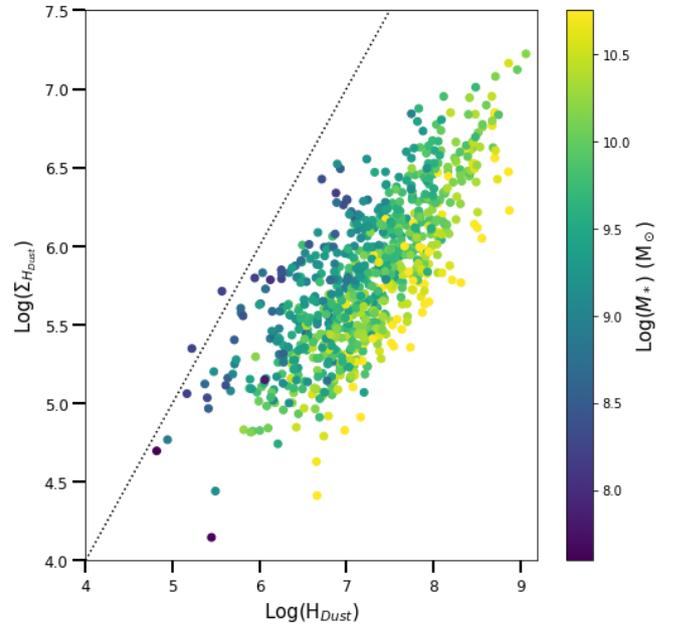

**Figure 7.** $\Sigma_{H_{\text{dust}}}$ as a function of $H_{\text{dust}}$, coloured by $M_*$. The dotted line is a 1:1 line. The correlation coefficient is 0.77.

Seeing as the dust mass and dust surface density definitions of $F$ and $H$ are correlated with one another, either may be used and yield similar results. To avoid repetition and eliminate redundancy, this analysis continues with only one set. As discussed above, $F_{\text{dust}}$ and $H_{\text{dust}}$ provide the more intuitive choice for this investigation. Thus, the remainder of this analysis focuses on $F_{\text{dust}}$ and $H_{\text{dust}}$.

Figure 8 presents the relationship between $F_{\text{dust}}$ and $H_{\text{dust}}$, showing a negative correlation, as expected from the way the two parameters are defined. There is a strong $M_*$ dependence visible, in the sense that the inverse correlation between $F_{dust}$ and $H_{dust}$ moves to higher values of $H_{dust}$ as $M_*$ increases. Galaxies with higher $F_{dust}$ tend to have lower values of $H_{dust}$. High $F_{dust}$ and low $H_{dust}$ correspond to a more foreground screen dust geometry with more optically thin dust. For galaxies with lower values of $F_{dust}$, the spread of $H_{dust}$ values increases. This indicates that as the geometry moves towards a greater proportion of distributed dust, there is greater variation in the optical depth observed. A numerical quantification of the anticorrelation between $F_{dust}$ and $H_{dust}$ is not especially illuminating, as this will be dependent on the specific choice of envelope line, which in turn will be dependent on the specific sample being used, and the determination of dust masses.

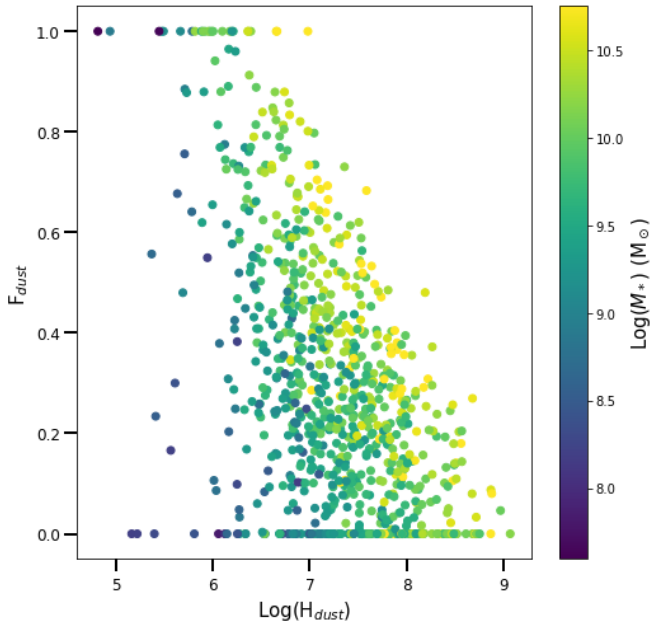

**Figure 8.** $F_{dust}$ as a function of $H_{dust}$, coloured by $M_*$. The correlation coefficient is -0.56.

We can use another approach to consider the qualitative geometry of the dust as well. If we assume the simplest scenario, that every galaxy has dust arranged in a foreground screen, that has an implication for how we can interpret $\Sigma_{M_d}$. This can be thought of as representing the thickness of the screen, if we consider the dust mass, $M_d$, as representative of the volume over which it is spread. In this scenario, we would expect that the BD would be correlated with $\Sigma_{M_d}$, since a thicker screen would produce a larger BD. We explore this explicitly in Figure 9, which shows $\Sigma_{M_d}$ as a function of BD, coloured by $H_{dust}$. We can see here that there is no correlation between $\Sigma_{M_d}$ and BD, with the possible exception of galaxies with the lowest $H_{dust}$. Galaxies with the highest $H_{dust}$ show the largest $\Sigma_{M_d}$ values, but are restricted to a narrow range of lowest BD. If the foreground screen interpretation is correct we would expect the thickest screen (largest $\Sigma_{M_d}$) to have the highest BD, but this is not the case. Accordingly, this implies that these galaxies do not favour a foreground screen geometry, and our interpretation of them as having dust that is mixed and distributed throughout the galaxy is more likely.

This permits low BD values to arise from the edges of the dust distribution, where low levels of obscuration can occur. The model proposed above (Figure 3) is supported by this result. Subsequently, we continue with our interpretation that galaxies with low BD, high $M_d$, and thus high $H_{dust}$, have a distributed dust geometry. This in turn is associated with dust that is more optically thick to Hα emission.

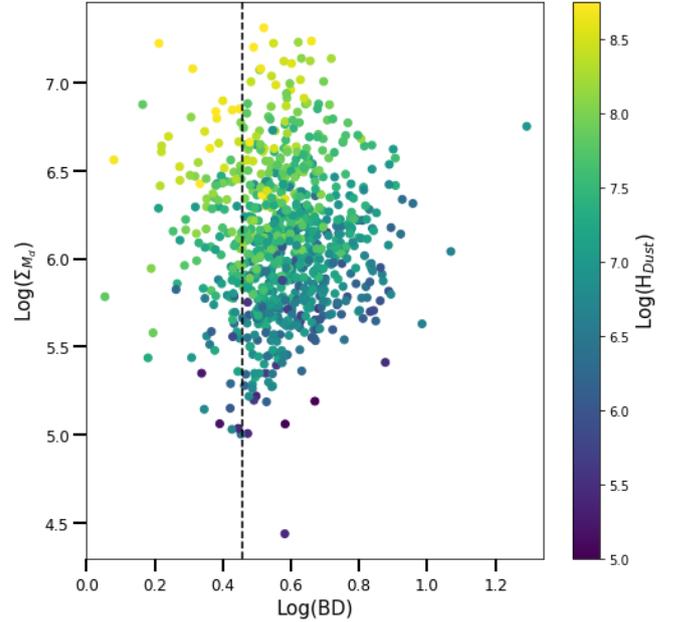

**Figure 9.** $\Sigma_{M_d}$ as a function of BD, coloured by $H_{dust}$. The dashed line represents the Case B value at BD = 2.86. The correlation coefficient is 0.013.

We now have two parameters which each provide a new way to quantify the optical depth and geometry of a galaxy's dust, with $F_{dust}$ being more directly related to the geometry and $H_{dust}$ being more directly related to the optical depth. The following analysis explores how these parameters are related to other observational quantities which are themselves expected to strongly trace the degree of optical depth of the dust.

## 5. Results

There is no correlation seen between $H_{dust}$ or $F_{dust}$ and $SFR_{Hα}/SFR_{FUV}$ (Figure 10). The notable difference between Figures 10a and 10b is that $H_{dust}$ shows a dependence on $M_*$, but $F_{dust}$ does not. This is a direct result of the way the two parameters are defined. $H_{dust}$ is defined such that it is strongly correlated to the $M_d$, and therefore also (indirectly) to the stellar mass (Figure 4a). $F_{dust}$ is defined in such a way that although it is mathematically dependent on the $M_d$ (through the BD envelope line), it is not strongly correlated with the $M_d$ (or $M_*$).

The lack of trend in Figures 10a and 10b indicates that the Hα and FUV are experiencing a similar degree of optically thick dust. This implies that the stars and the gas are closely co-located, with the optically thick effects impacting both to the same degree. If this were not the case, and the emission from the stars and gas were experiencing different levels of obscuration, then Figure 10a would show a positive correlation

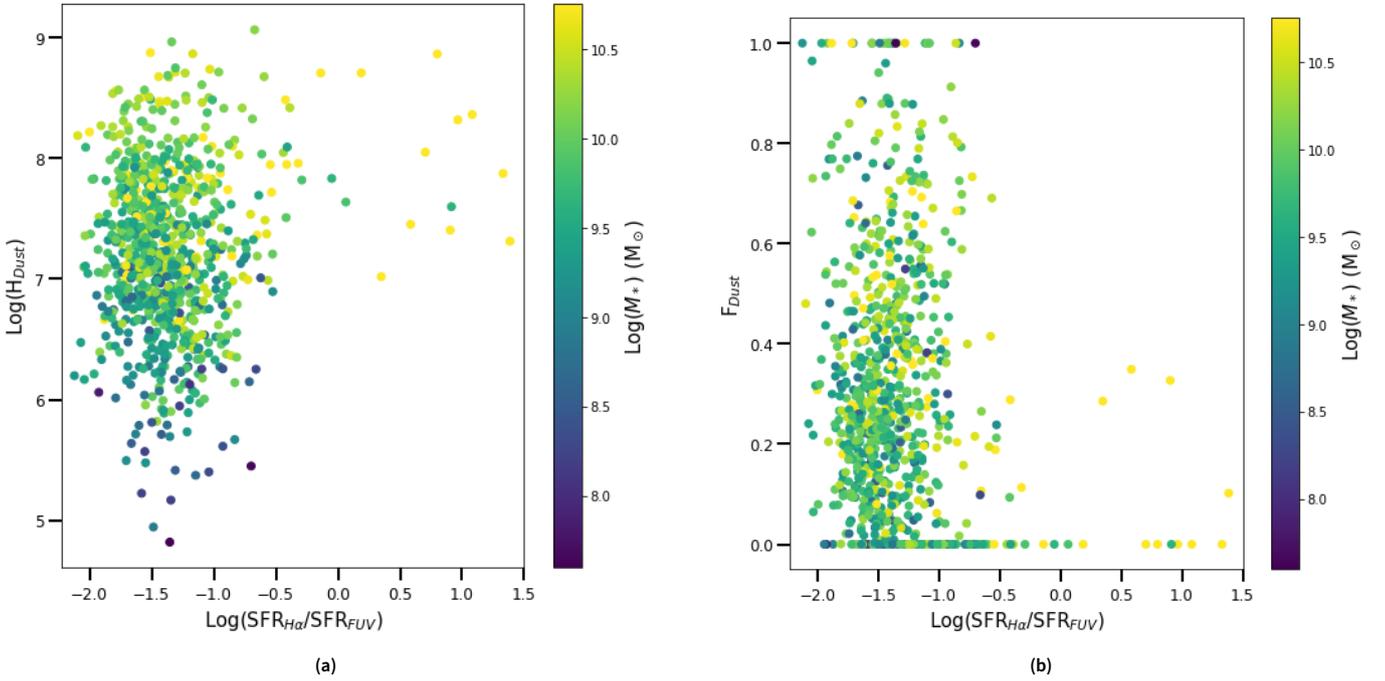

**Figure 10.** The relationships between SFR$_{H\alpha}$/SFR$_{FUV}$ and (a) $H_{dust}$, and (b) $F_{dust}$, each coloured by $M_*$. The correlation coefficient for panel (a) is 0.093 and the correlation coefficient for panel (b) is -0.14.

between $H_{dust}$ and SFR$_{H\alpha}$/SFR$_{FUV}$, while Figure 10b would show a negative trend.

To further explore the relationships between $H_{dust}$, $F_{dust}$, and the H$\alpha$ deficit, we define four independent mass-limited redshift bins. The full sample spans a redshift range $0 < z < 0.35$. The redshift bins were defined as shown in Figure 11 and Table 3.

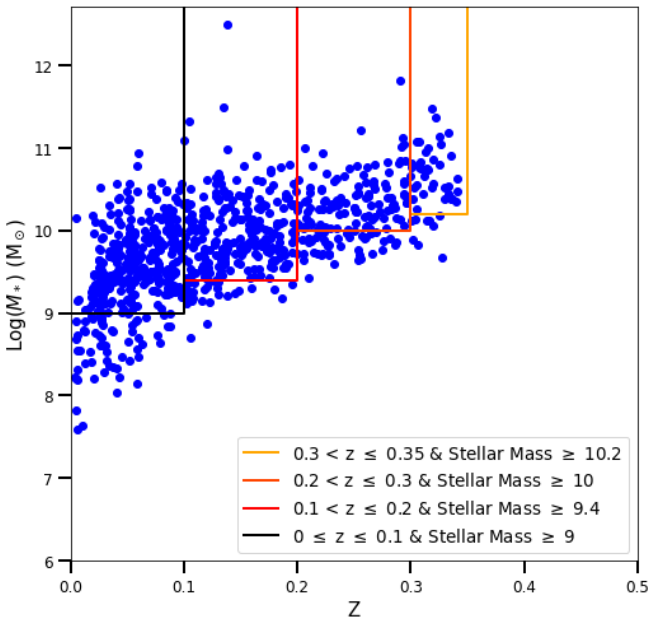

**Figure 11.** The four volume-limited samples used to explore any redshift and mass dependencies.

Figures 12a and 12b show the relationships between $H_{dust}$

**Table 3.** The number of objects in each of the mass-limited redshift bins.

| Redshift | Number of objects | Mass limit | Objects below mass limit |
|---|---|---|---|
| $0.0 \leq z \leq 0.1$ | 321 | 9 | 63 |
| $0.1 \leq z \leq 0.2$ | 225 | 9.4 | 35 |
| $0.2 \leq z \leq 0.3$ | 109 | 10 | 41 |
| $0.3 \leq z \leq 0.35$ | 39 | 10.2 | 9 |

and $F_{dust}$ with the H$\alpha$ deficit. $H_{dust}$ has a positive correlation with the H$\alpha$ deficit, whereas $F_{dust}$ has a negative correlation. These opposing trends are expected due to the negative correlation between $H_{dust}$ and $F_{dust}$ (Figure 8). Higher H$\alpha$ deficit values reflect the greater optical depth experienced by the H$\alpha$ compared to the FIR, and are seen to correspond to high values of $H_{dust}$. Conversely, and self-consistently, this leads to lower values of $F_{dust}$, related to the distributed dust geometry. This figure confirms that the dust properties are quantified cleanly through $H_{dust}$ and $F_{dust}$. A higher H$\alpha$ deficit corresponds to higher $H_{dust}$ values (increased optical depth), and low values of $F_{dust}$ (distributed dust geometry). The correlation between $H_{dust}$ and the H$\alpha$ deficit is not caused by the presence of the BD in the calculation of both $H_{dust}$ and the corrected H$\alpha$ luminosities. This correlation is still observed if the uncorrected H$\alpha$ luminosities are used.

Figure 13 shows the relationship between $F_{dust}$ and $H_{dust}$ with the H$\alpha$ deficit for each of the four mass-limited redshift bins. The line in Figure 12a is not a fit, simply defined to guide the eye. This same line is overlayed onto each of the mass-limited redshift bins in Figure 13a. In each of these bins, while the sample becomes progressively limited to the higher $M_*$ systems, the trend is consistent with the full sample. This demonstrates that the broad relationship between $H_{dust}$ and the Hα deficit does not evolve over this redshift range.

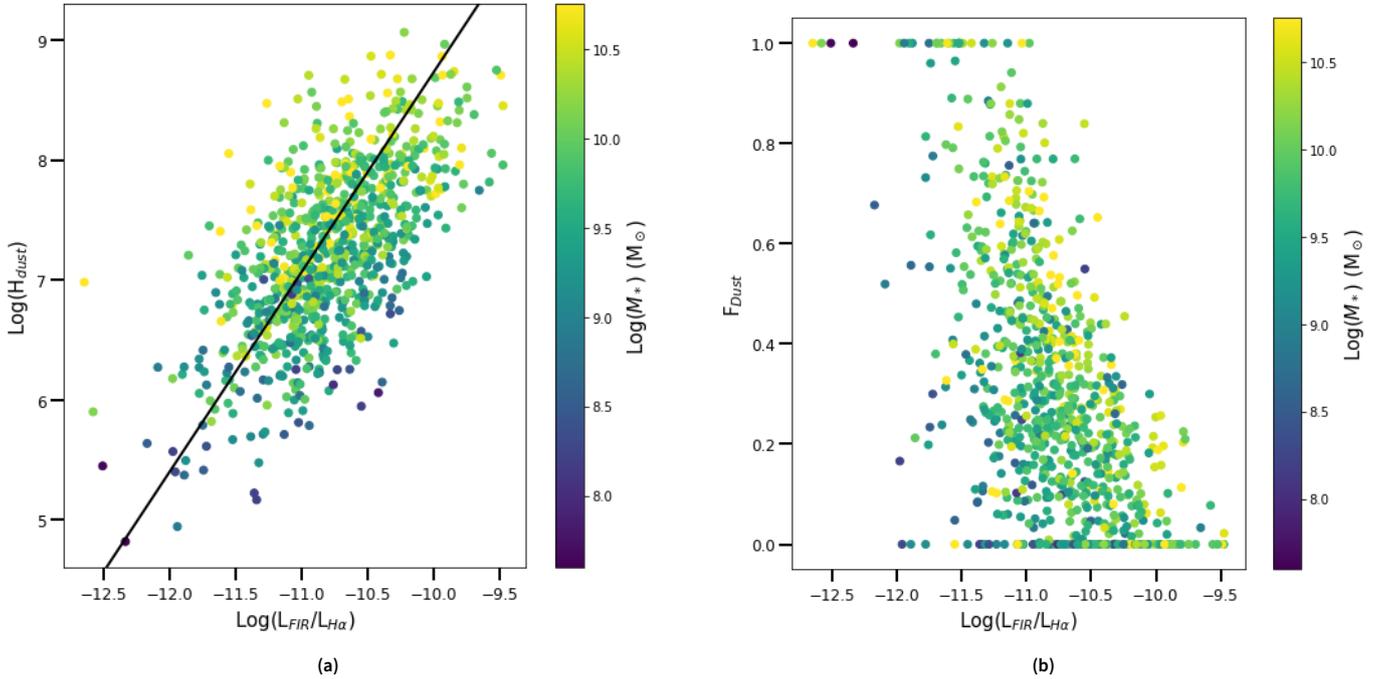

**Figure 12.** (a) $H_{dust}$ and (b) $F_{dust}$ as a function of Hα deficit coloured by $M_*$. The line in (a) is not a fit, simply given to guide the eye. The correlation coefficient for panel (a) is 0.64 and the correlation coefficient for panel (b) is -0.59.

Figure 13b shows the relationship between $F_{dust}$ and the Hα deficit over the four redshift bins. Since each redshift bin samples galaxies limited to higher $M_*$, the mass dependence visible in Figure 12b is highlighted. As redshift increases, higher $M_*$ systems are selected. Accordingly, these correspond to galaxies with more distributed dust geometries and more optically thick dust. This is a consequence of selecting higher $M_*$ galaxies at higher $z$, however, not any intrinsic evolutionary effect.

We also explore the ratio between the FIR luminosity and the FUV luminosity which, analogous to the Hα deficit, will be referred to here as the "FUV deficit". Figure 14 shows the relationships between $H_{dust}$ and $F_{dust}$ with the FUV deficit. The trends seen in this Figure reiterate those seen in the corresponding Hα deficit figures. Once again, higher FUV deficit values indicate a greater optical depth for the shorter wavelength light, corresponding to higher $H_{dust}$ values and the distributed dust geometry of low values of $F_{dust}$.

## 6.  Discussion

We have established that $H_{dust}$ and $F_{dust}$ appear to be sensitive to short wavelength deficits, which we attribute to a degree of optically thick obscuration. We can now use that result to explore how the dust geometry as traced by $H_{dust}$ and $F_{dust}$ varies for different galaxy properties. We look specifically at the relationship with galaxy stellar mass, SFR, and specific SFR.

Figure 15 illustrates how $H_{dust}$ and $F_{dust}$ vary with $M_*$.

Figure 15a shows $H_{dust}$ increasing with stellar mass, which is to be expected given its strong link to the dust mass, and the correlation between dust mass and stellar mass. The SFR also clearly correlates with stellar mass, and, for a given SFR, $H_{dust}$ and stellar mass are strongly correlated. This implies that the most optically thick dust, at the highest values of $H_{dust}$, are, as might be expected, in the systems with both the highest stellar mass and the highest SFRs. The most extreme such systems (in the upper right, with the highest $M_*$) are likely absent from this diagram. This is due to the selection bias introduced by complete optically thick obscuration of the Balmer lines in the highest dust mass galaxies (corresponding to the absence of data in the upper right of Figure 2a).

Figure 15b shows the relationship between $F_{dust}$ and stellar mass coloured by $SFR_{Hα}$. This figure shows a slight increase in the median $F_{dust}$ values with $M_*$. The lack of a strong correlation between $F_{dust}$ and $M_*$ suggests that the proportion of foreground screen or distributed geometry present in a galaxy is not primarily influenced by the galaxy's stellar mass. It can also be seen that at a given value of $SFR_{Hα}$, $F_{dust}$ decreases as $M_*$ increases (a slightly inverse correlation at fixed SFR). This indicates that for a given SFR, a lower mass galaxy will favour a more foreground screen geometry, while a higher mass galaxy will show a more distributed dust geometry.

Figure 16 presents the dependence of $H_{dust}$ and $F_{dust}$ on $SFR_{Hα}$. Figure 16a shows a positive trend between $H_{dust}$ and $SFR_{Hα}$, which then flattens and turns downwards at the highest values of $SFR_{Hα}$. This is attributable to the absence of the highest $H_{dust}$ values. Again, this absence of data points in the high $H_{dust}$, high SFR, high $M_*$ region can be attributed to the selection bias from the complete obscuration of the Balmer lines. At a given $M_*$, $H_{dust}$ decreases with increasing $SFR_{H\alpha}$. This indicates that for galaxies of a given mass, higher SFRs occur in those with less optically thick dust.

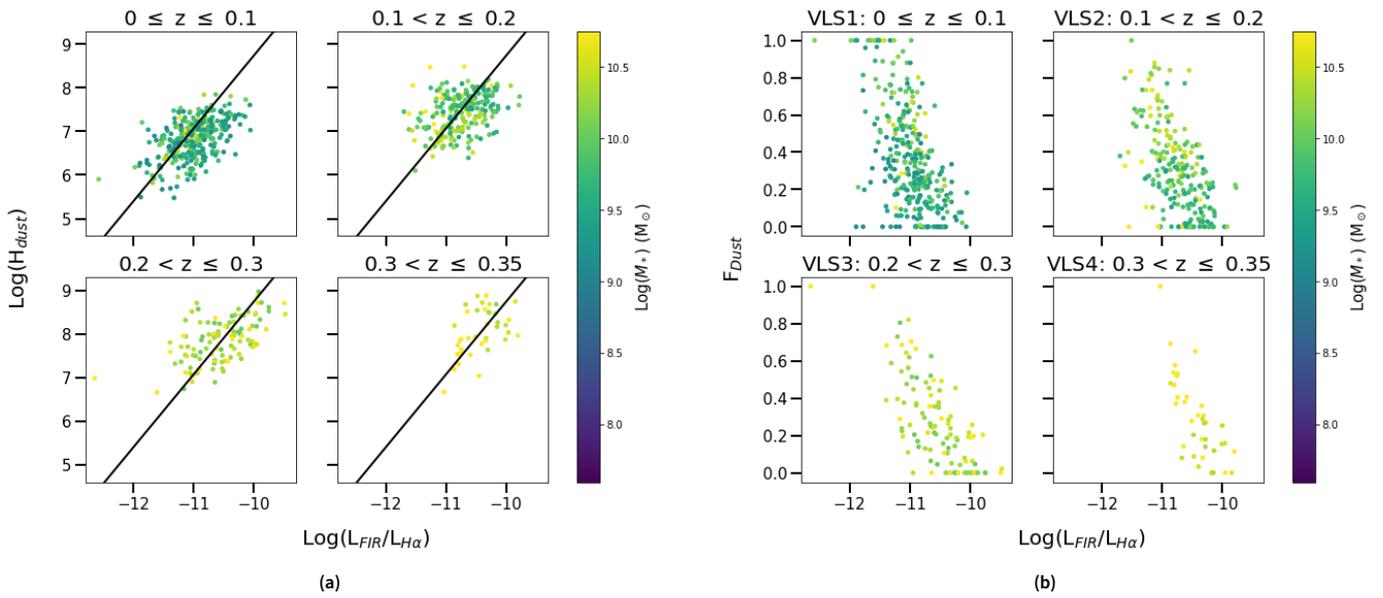

**Figure 13.** (a) $H_{\rm dust}$ as a function of H$\alpha$ deficit coloured by $M_*$ for the mass-limited redshift bins, and (b) $F_{\rm dust}$ as a function of H$\alpha$ deficit coloured by $M_*$ for the mass-limited redshift bins. The lines in (a) are the same as in Figure 12a to guide the eye, and highlight that, while the masses sampled in higher redshift bins increase, the galaxy population follows the same trend.

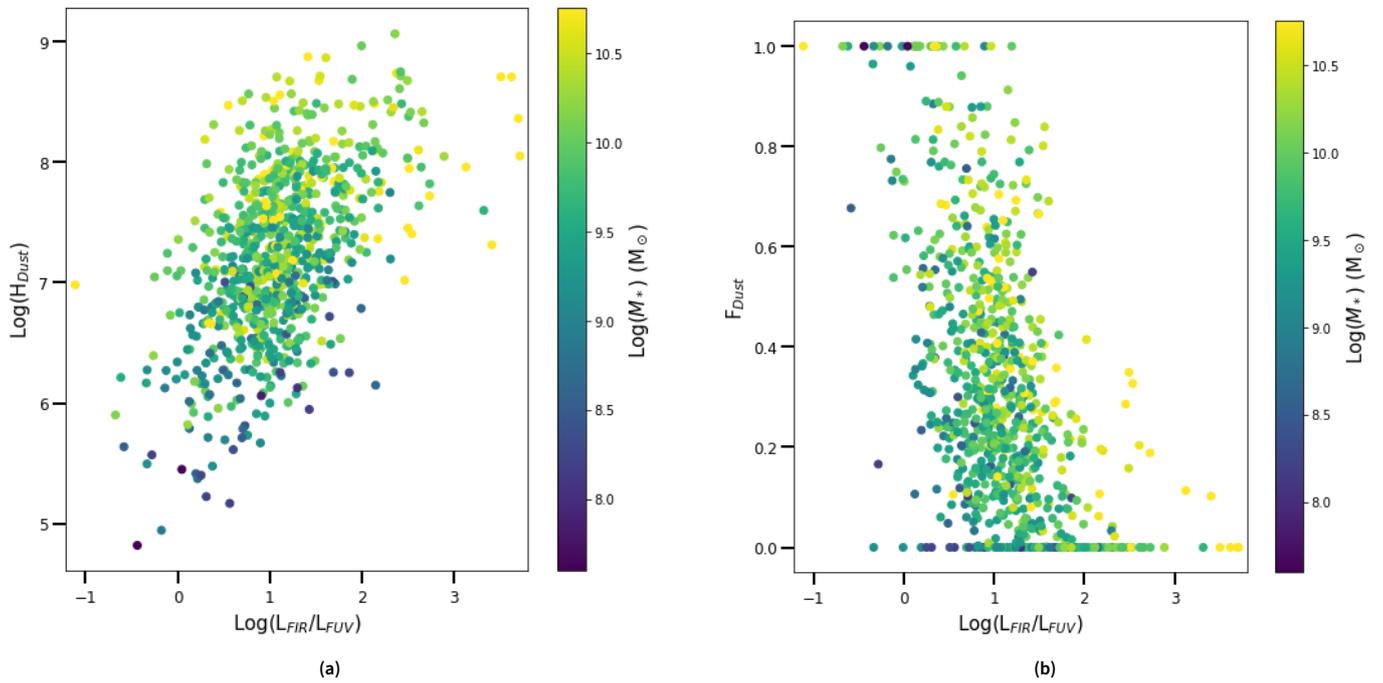

**Figure 14.** (a) $H_{\rm dust}$ as a function of FUV deficit coloured by $M_*$, and (b) $F_{\rm dust}$ as a function of FUV deficit coloured by $M_*$. The correlation coefficient for panel (a) is 0.52 and the correlation coefficient for panel (b) is -0.52.

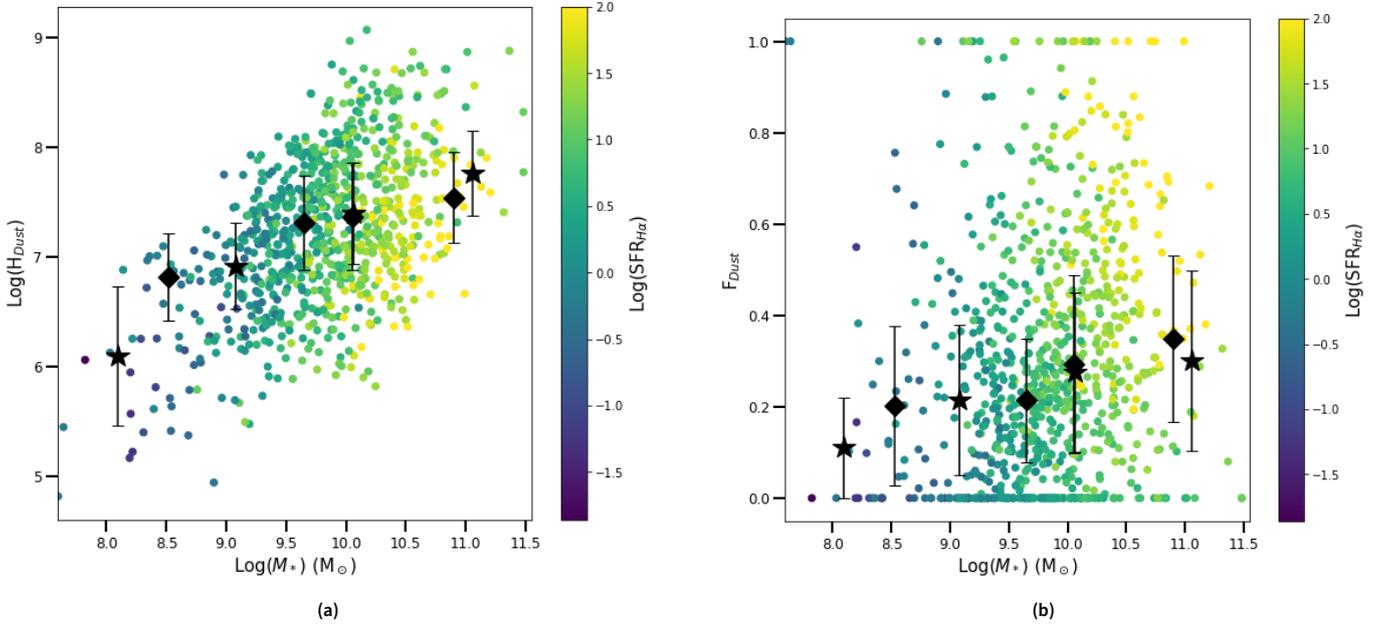

**Figure 15.** The relationship between $M_*$ and (a) $H_{dust}$, and (b) $F_{dust}$ coloured by $SFR_{H\alpha}$. The data are separated into four $M_*$ bins. The black stars represent the median value in each bin when the bins are evenly spaced. The black diamonds represent the median value in each bin when there are approximately the same number of objects in each bin. The errorbars show the median absolute deviations. The correlation coefficient for panel (a) is 0.50 and the correlation coefficient for panel (b) is 0.13.

$F_{dust}$ is clearly correlated with $SFR_{H\alpha}$, as seen in Figure 16b. The influence of $M_*$ in this figure can also be seen. The strong $M_*$-$SFR_{H\alpha}$ relationship is visible, but equally clear is that for a fixed $M_*$ a strong positive correlation exists between $F_{dust}$ and $SFR_{H\alpha}$. This indicates that the proportion of foreground screen contribution to the dust geometry increases as $SFR_{H\alpha}$ increases.

The negative correlation between $H_{dust}$ and $F_{dust}$ can be seen in Figure 16 in the $M_*$ dependence in each panel. In both the broad trend between $M_*$ and SFR is apparent. However, in 16a, at a fixed $M_*$ $H_{dust}$ decreases with $SFR_{H\alpha}$, while in 16b at a fixed $M_*$ $F_{dust}$ increases with $SFR_{H\alpha}$.

Figure 17 rounds out the collection by showing how $H_{dust}$ and $F_{dust}$ depend on $sSFR_{H\alpha}$. Figure 17a shows the relationship for $H_{dust}$. Here, the median values show $H_{dust}$ decreasing as $sSFR$ increases. This trend of $H_{dust}$ decreasing as $sSFR$ increases is consistent with Figure 16a in which, for a given $M_*$, $H_{dust}$ decreases with increasing $SFR_{H\alpha}$. The $M_*$ colour coding shows how the use of $sSFR$ has removed much of the stellar mass dependence seen in Figure 16a.

Figure 17b shows the correlation between $F_{dust}$ and $sSFR_{H\alpha}$. The use of $sSFR$ removes the $M_*$ dependence visible in Figure 16b, demonstrating that the contribution of foreground screen dust to the geometry clearly increases with an increase in H$\alpha$ star formation intensity. Galaxies with high $sSFR$ and high $F_{dust}$ can clearly be associated with the maximal foreground screen dust geometry. Conversely, galaxies with low $sSFR$ correspond to a more distributed dust geometry.

Taken together, these results can be interpreted as confirming that galaxies with high SFR and sSFR are best characterised with a "maximal foreground screen" dust geometry. This is consistent with the widely adopted choice of dust model for starburst galaxies (e.g., Calzetti et al. 1997; Calzetti et al. 2000; Calzetti 2001a). Conversely, galaxies with low SFR and sSFR appear to strongly favour the "maximal distributed dust" geometry. They also demonstrated the least optically thick dust, evidenced in their low values of $H_{dust}$, while the high SFR and high mass systems contain a greater degree of optically thick dust. The systems with the highest values of $H_{dust}$ are also those with the highest masses and lowest sSFR.

Ahmed et al. (in preparation) continue to explore the use of $F_{dust}$ and $H_{dust}$ by incorporating radio luminosities in their analysis. The use of the longer wavelength radio luminosity, which is not sensitive to obscuration, allows for an independent comparison to the results seen here when using the H$\alpha$ luminosity. Comparison of the trends seen when using a wavelength that is susceptible to obscuration with one that is not will help illuminate which trends observed in this paper, if any, may be a consequence of the remaining loss of H$\alpha$ emission due to obscuration.

## 7. Conclusion

We have introduced two new parameters, $H_{dust}$ and $F_{dust}$ that provide a novel method to quantify the properties of obscuration in galaxies. These parameters both use the BD and dust mass combined to infer more information about dust geometry and optical depth. We explored the use of dust surface density in place of dust mass in these parameters. Dust mass was demonstrated to be the more robust and effective approach in the definition of these new dust parameters. We have demonstrated that the BD and dust mass are explicitly linked to the degree of optically thick obscuration by comparison to parameters quantifying an Hα or UV deficit ($L_{FIR}/L_{H\alpha}$ or $L_{FIR}/L_{FUV}$). We show that $H_{dust}$ is a good tracer for the degree of optically thick dust in a galaxy, and that $F_{dust}$ is an efficient parameter to quantify the form of the dust geometry.

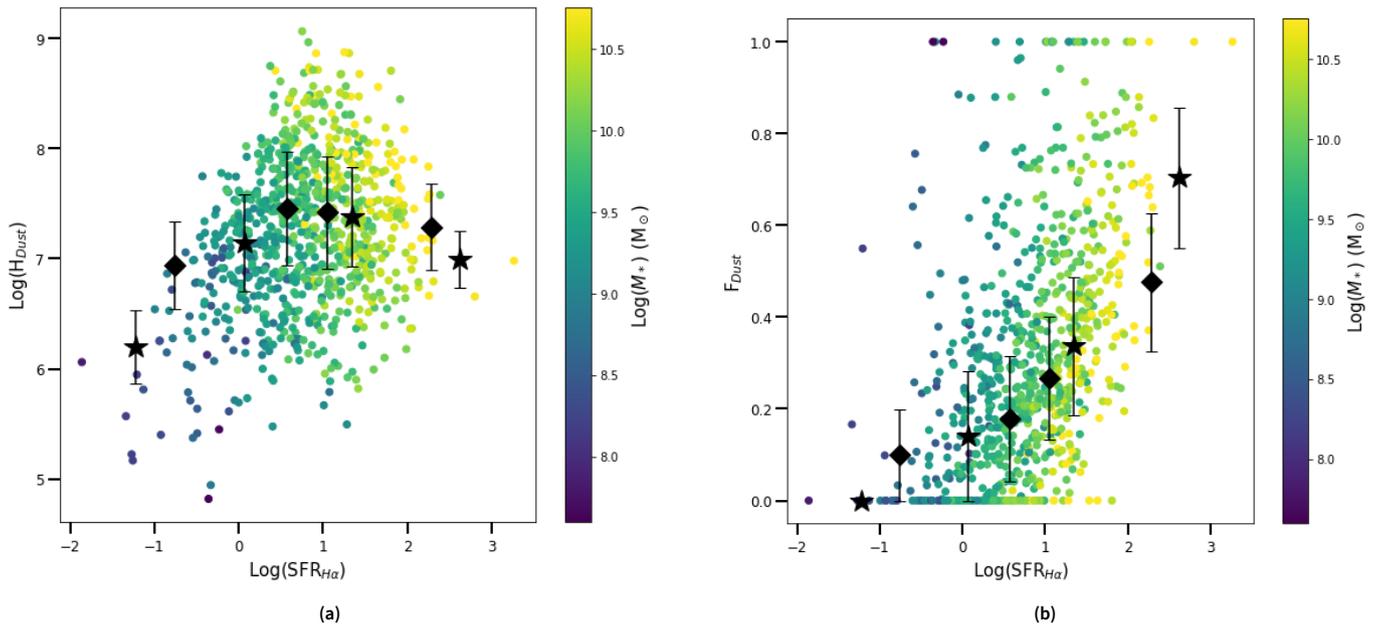

**Figure 16.** The relationship between SFR$_{H\alpha}$ and (a) $H_{dust}$, and (b) $F_{dust}$, coloured by $M_*$. The data are separated into four SFR bins. The black stars represent the median value in each bin when the bins are evenly spaced. The black diamonds represent the median value in each bin when there are approximately the same number of objects in each bin. The errorbars show the median absolute deviations. If the errorbars are not visible, it is because they are smaller than the star markers. The correlation coefficient for panel (a) is 0.26 and the correlation coefficient for panel (b) is 0.49.

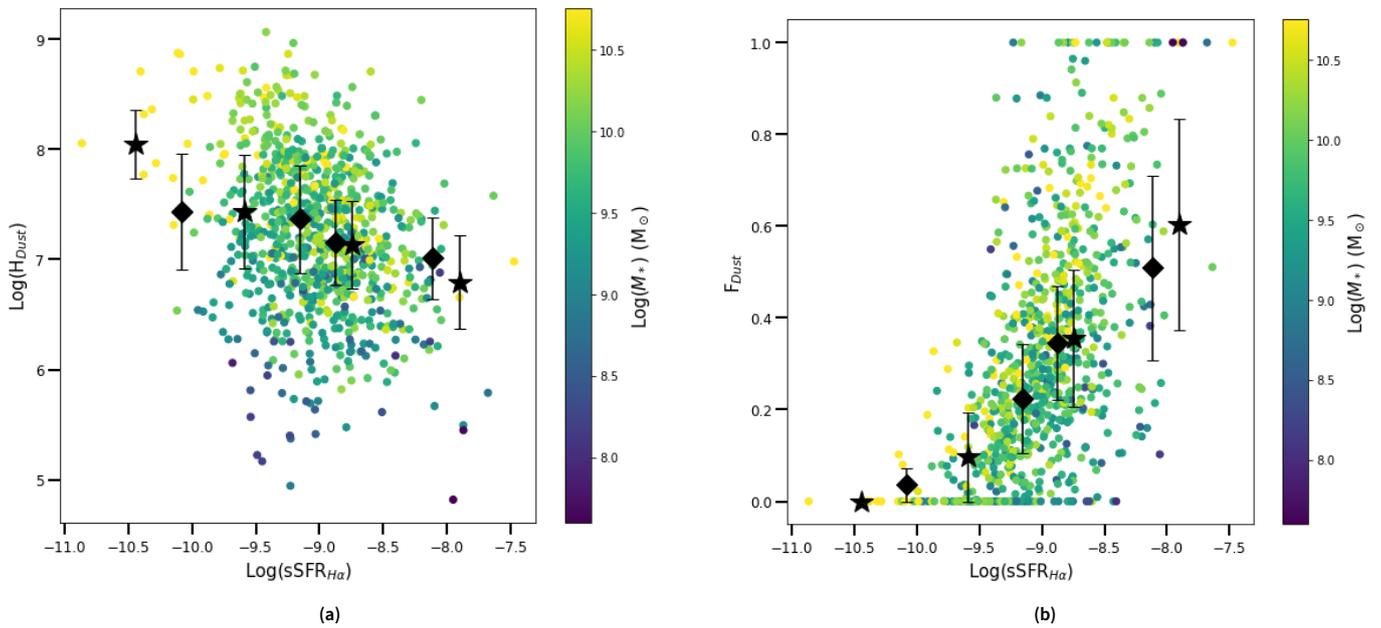

**Figure 17.** The relationship between sSFR$_{H\alpha}$ and (a) $H_{dust}$, and (b) $F_{dust}$, coloured by $M_*$. The data are separated into four sSFR bins. The black stars represent the median value in each bin when the bins are evenly spaced. The black diamonds represent the median value in each bin when there are approximately the same number of objects in each bin. The errorbars show the median absolute deviations. If the errorbars are not visible, it is because they are smaller than the star markers. The correlation coefficient for panel (a) is -0.28 and the correlation coefficient for panel (b) 0.596.

We use these parameters to explore how dust opacity and geometry relate to other basic galaxy properties. We see that $F_{dust}$ has no dependence on stellar mass, due to the nature of how it is defined, but $H_{dust}$ does correlate with $M_*$ (Figure 15), as a consequence of its link with dust mass by definition. However, its ability to quantify the optical depth is not stellar mass dependent, and the range of optical depth sampled (the range of $H_{dust}$ values) becomes larger at higher $M_*$.

We have also shown that these parameters are linked to SFR and sSFR. It is clear that the assumption of a maximal foreground screen dust model is appropriate for highly star forming galaxies, when measuring obscuration-sensitive properties such as Hα (Figures 16b and 17b). It is also the case that the highest SFR galaxies show the greatest level of optical depth (Figure 16a), even while these may be low sSFR systems given their large $M_*$ (Figure 17a). Together this implies that in the highest SFR systems there is likely to be a deficit in the measured Hα, even after obscuration correction, due to optically thick dust.

The ability to measure dust geometry and have a quantitative constraint on the degree of optically thick dust opens many possibilities for better measurement and understanding of fundamental galaxy properties. Future work will investigate dependencies on metallicity, inclination, morphology, environment, redshift, and other galaxy properties. Quantifying a series of models encompassing all geometries may also provide a fruitful direction for future work. We will also explore whether there are further refinements in how we may define $H_{dust}$ and $F_{dust}$, or incorporate other properties to provide still better insights into the dust properties of galaxies.

### Data Availability
The GAMA data are all publicly available[a].

### Acknowledgement
GAMA is a joint European–Australasian project based around a spectroscopic campaign using the Anglo-Australian Telescope. The GAMA input catalogue is based on data taken from the Sloan Digital Sky Survey and the UKIRT Infrared Deep Sky Survey. Complementary imaging of the GAMA regions is being obtained by a number of independent survey programmes including GALEX MIS, VST KiDS, VISTA VIKING, WISE, Herschel-ATLAS, GMRT and ASKAP providing UV to radio coverage. GAMA is funded by the STFC (UK), the ARC (Australia), the AAO, and the participating institutions. The GAMA website is http://www.gama-survey.org/ . Based on observations made with ESO Telescopes at the La Silla Paranal Observatory under programme IDs 179.A-2004 and 177.A-3016.

### References

Baldry, I K, J Liske, M J I Brown, A S G Robotham, S P Driver, L Dunne, M Alpaslan, et al. 2018. Galaxy And Mass Assembly: the G02 field, Herschel–ATLAS target selection and data release 3 [in en]. *Monthly Notices of the Royal Astronomical Society* 474, no. 3 (March): 3875–3888. ISSN: 0035-8711, 1365-2966, accessed March 14, 2023. https://doi.org/10.1093/mnras/stx3042. http://academic.oup.com/mnras/article/474/3/3875/4657188.

Baldwin, J. A., M. M. Phillips, and R. Terlevich. 1981. Classification parameters for the emission-line spectra of extragalactic objects [in en]. *Publications of the Astronomical Society of the Pacific* 93 (February): 5. ISSN: 0004-6280, 1538-3873, accessed August 7, 2023. https://doi.org/10.1086/130766. http://iopscience.iop.org/article/10.1086/130766.

Bellstedt, Sabine, Simon P Driver, Aaron S G Robotham, Luke J M Davies, Kamran R J Bogue, Robin H W Cook, Abdolhosein Hashemizadeh, et al. 2020a. Galaxy And Mass Assembly (GAMA): assimilation of KiDS into the GAMA database [in en]. *Monthly Notices of the Royal Astronomical Society* 496, no. 3 (August): 3235–3256. ISSN: 0035-8711, 1365-2966, accessed August 22, 2024. https://doi.org/10.1093/mnras/staa1466. https://academic.oup.com/mnras/article/496/3/3235/5850776.

———. 2020b. Galaxy And Mass Assembly (GAMA): assimilation of KiDS into the GAMA database. *Monthly Notices of the Royal Astronomical Society* 496, no. 3 (August): 3235–3256. https://doi.org/10.1093/mnras/staa1466. arXiv: 2005.11215 [astro-ph.GA].

Bruzual, G., and S. Charlot. 2003. Stellar population synthesis at the resolution of 2003 [in en]. *Monthly Notices of the Royal Astronomical Society* 344, no. 4 (October): 1000–1028. ISSN: 0035-8711, 1365-2966, accessed August 30, 2024. https://doi.org/10.1046/j.1365-8711.2003.06897.x. https://academic.oup.com/mnras/article/344/4/1000/968846.

Buat, V., J. Iglesias-Páramo, M. Seibert, D. Burgarella, S. Charlot, D. C. Martin, C. K. Xu, et al. 2005. Dust Attenuation in the Nearby Universe: A Comparison between Galaxies Selected in the Ultraviolet and in the Far-Infrared [in en]. Publisher: IOP Publishing, *The Astrophysical Journal* 619, no. 1 (January): L51. ISSN: 0004-637X, accessed August 22, 2024. https://doi.org/10.1086/423241. https://iopscience.iop.org/article/10.1086/423241/meta.

Calura, F., F. Pozzi, G. Cresci, P. Santini, C. Gruppioni, L. Pozzetti, R. Gilli, F. Matteucci, and R. Maiolino. 2017. The dust-to-stellar mass ratio as a valuable tool to probe the evolution of local and distant star-forming galaxies [in en]. *Monthly Notices of the Royal Astronomical Society* 465, no. 1 (February): 54–67. ISSN: 0035-8711, 1365-2966, accessed April 2, 2025. https://doi.org/10.1093/mnras/stw2749. https://academic.oup.com/mnras/article-lookup/doi/10.1093/mnras/stw2749.

Calzetti, Daniela. 2001a. The Dust Opacity of Star-forming Galaxies [in en]. *Publications of the Astronomical Society of the Pacific* 113, no. 790 (December): 1449–1485. ISSN: 0004-6280, 1538-3873, accessed November 8, 2023. https://doi.org/10.1086/324269. http://iopscience.iop.org/article/10.1086/324269.

———. 2001b. The effects of dust on the spectral energy distribution of star-forming galaxies [in en]. *New Astronomy Reviews* 45, nos. 9-10 (October): 601–607. ISSN: 13876473, accessed February 27, 2023. https://doi.org/10.1016/S1387-6473(01)00144-0. https://linkinghub.elsevier.com/retrieve/pii/S1387647301001440.

Calzetti, Daniela, Lee Armus, Ralph C. Bohlin, Anne L. Kinney, Jan Koornneef, and Thaisa Storchi-Bergmann. 2000. The Dust Content and Opacity of Actively Star-forming Galaxies [in en]. *The Astrophysical Journal* 533, no. 2 (April): 682–695. ISSN: 0004-637X, 1538-4357, accessed November 24, 2023. https://doi.org/10.1086/308692. https://iopscience.iop.org/article/10.1086/308692.

Calzetti, Daniela, Anne L. Kinney, and Thaisa Storchi-Bergmann. 1994. Dust extinction of the stellar continua in starburst galaxies: The ultraviolet and optical extinction law [in en]. *The Astrophysical Journal* 429 (July): 582. ISSN: 0004-637X, 1538-4357, accessed November 27, 2023. https://doi.org/10.1086/174346. http://adsabs.harvard.edu/doi/10.1086/174346.

Calzetti, Daniela, Gerhardt R. Meurer, Ralph C. Bohlin, Donald R. Garnett, Anne L. Kinney, Claus Leitherer, and Thaisa Storchi-Bergmann. 1997. Dust and Recent Star Formation in the Core of NGC 5253 [in en]. *The Astronomical Journal* 114 (November): 1834. ISSN: 00046256, accessed November 27, 2023. https://doi.org/10.1086/118609. http://adsabs.harvard.edu/cgi-bin/bib_query?1997AJ....114.1834C.

Casasola, V., S. Bianchi, P. De Vis, L. Magrini, E. Corbelli, C. J. R. Clark, J. Fritz, et al. 2020. The ISM scaling relations in DustPedia late-type galaxies: A benchmark study for the Local Universe [in en]. *Astronomy & Astrophysics* 633 (January): A100. ISSN: 0004-6361, 1432-0746, accessed April 2, 2025. https://doi.org/10.1051/0004-6361/201936665. https://www.aanda.org/10.1051/0004-6361/201936665.

Casasola, Viviana, Simone Bianchi, Laura Magrini, Aleksandr V. Mosenkov, Francesco Salvestrini, Maarten Baes, Francesco Calura, et al. 2022. The resolved scaling relations in DustPedia: Zooming in on the local Universe [in en]. *Astronomy & Astrophysics* 668 (December): A130. ISSN: 0004-6361, 1432-0746, accessed April 2, 2025. https://doi.org/10.1051/0004-6361/202245043. https://www.aanda.org/10.1051/0004-6361/202245043.

Charlot, Stéphane, and S. Michael Fall. 2000. A Simple Model for the Absorption of Starlight by Dust in Galaxies. *The Astrophysical Journal* 539, no. 2 (August): 718–731. https://doi.org/10.1086/309250. arXiv: astro-ph/0003128 [astro-ph].

Clemens, M. S., M. Negrello, G. De Zotti, J. Gonzalez-Nuevo, L. Bonavera, G. Cosco, G. Guarese, et al. 2013. Dust and star formation properties of a complete sample of local galaxies drawn from the Planck Early Release Compact Source Catalogue [in en]. *Monthly Notices of the Royal Astronomical Society* 433, no. 1 (July): 695–711. ISSN: 0035-8711, 1365-2966, accessed July 10, 2023. https://doi.org/10.1093/mnras/stt760. http://academic.oup.com/mnras/article/433/1/695/1039514/Dust-and-star-formation-properties-of-a-complete.

Cortese, L., L. Ciesla, A. Boselli, S. Bianchi, H. Gomez, M. W. L. Smith, G. J. Bendo, et al. 2012. The dust scaling relations of the *Herschel* Reference Survey [in en]. *Astronomy & Astrophysics* 540 (April): A52. ISSN: 0004-6361, 1432-0746, accessed July 13, 2023. https://doi.org/10.1051/0004-6361/201118499. http://www.aanda.org/10.1051/0004-6361/201118499.

Cunha, Elisabete da, Stéphane Charlot, and David Elbaz. 2008. A simple model to interpret the ultraviolet, optical and infrared emission from galaxies [in en]. *Monthly Notices of the Royal Astronomical Society* 388, no. 4 (August): 1595–1617. ISSN: 00358711, 13652966, accessed July 10, 2023. https://doi.org/10.1111/j.1365-2966.2008.13535.x. https://academic.oup.com/mnras/article-lookup/doi/10.1111/j.1365-2966.2008.13535.x.

De Vis, P., L. Dunne, S. Maddox, H. L. Gomez, C. J. R. Clark, A. E. Bauer, S. Viaene, et al. 2017. *Herschel* -ATLAS: revealing dust build-up and decline across gas, dust and stellar mass selected samples – I. Scaling relations [in en]. *Monthly Notices of the Royal Astronomical Society* 464, no. 4 (February): 4680–4705. ISSN: 0035-8711, 1365-2966, accessed April 2, 2025. https://doi.org/10.1093/mnras/stw2501. https://academic.oup.com/mnras/article-lookup/doi/10.1093/mnras/stw2501.

Driver, S. P., D. T. Hill, L. S. Kelvin, A. S. G. Robotham, J. Liske, P. Norberg, I. K. Baldry, et al. 2011. Galaxy and Mass Assembly (GAMA): survey diagnostics and core data release: GAMA [in en]. *Monthly Notices of the Royal Astronomical Society* 413, no. 2 (May): 971–995. ISSN: 00358711, accessed January 18, 2023. https://doi.org/10.1111/j.1365-2966.2010.18188.x. https://academic.oup.com/mnras/article-lookup/doi/10.1111/j.1365-2966.2010.18188.x.

Driver, S. P., C. C. Popescu, R. J. Tuffs, J. Liske, A. W. Graham, P. D. Allen, and R. De Propris. 2007. The Millennium Galaxy Catalogue: the B-band attenuation of bulge and disc light and the implied cosmic dust and stellar mass densities [in en]. *Monthly Notices of the Royal Astronomical Society* 379, no. 3 (August): 1022–1036. ISSN: 0035-8711, 1365-2966, accessed July 10, 2023. https://doi.org/10.1111/j.1365-2966.2007.11862.x. https://academic.oup.com/mnras/article-lookup/doi/10.1111/j.1365-2966.2007.11862.x.

Driver, Simon P, Sabine Bellstedt, Aaron S G Robotham, Ivan K Baldry, Luke J Davies, Jochen Liske, Danail Obreschkow, et al. 2022. Galaxy And Mass Assembly (GAMA): Data Release 4 and the z < 0.1 total and z < 0.08 morphological galaxy stellar mass functions [in en]. *Monthly Notices of the Royal Astronomical Society* 513, no. 1 (April): 439–467. ISSN: 0035-8711, 1365-2966, accessed March 14, 2023. https://doi.org/10.1093/mnras/stac472. https://academic.oup.com/mnras/article/513/1/439/6540978.

Driver, Simon P., Angus H. Wright, Stephen K. Andrews, Luke J. Davies, Prajwal R. Kafle, Rebecca Lange, Amanda J. Moffett, et al. 2016. Galaxy And Mass Assembly (GAMA): Panchromatic Data Release (far-UV–far-IR) and the low- z energy budget [in en]. *Monthly Notices of the Royal Astronomical Society* 455, no. 4 (February): 3911–3942. ISSN: 0035-8711, 1365-2966, accessed March 14, 2023. https://doi.org/10.1093/mnras/stv2505. https://academic.oup.com/mnras/article-lookup/doi/10.1093/mnras/stv2505.

Gordon, Yjan A., Matt S. Owers, Kevin A. Pimbblet, Scott M. Croom, Mehmet Alpaslan, Ivan K. Baldry, Sarah Brough, et al. 2017. Galaxy and Mass Assembly (GAMA): active galactic nuclei in pairs of galaxies. *Monthly Notices of the Royal Astronomical Society* 465, no. 3 (March): 2671–2686. https://doi.org/10.1093/mnras/stw2925. arXiv: 1611.03376 [astro-ph.GA].

Groves, Brent, Jarle Brinchmann, and Carl Jakob Walcher. 2012. The Balmer decrement of Sloan Digital Sky Survey galaxies: The Balmer decrement of SDSS galaxies [in en]. *Monthly Notices of the Royal Astronomical Society* 419, no. 2 (January): 1402–1412. ISSN: 00358711, accessed June 12, 2023. https://doi.org/10.1111/j.1365-2966.2011.19796.x. https://academic.oup.com/mnras/article-lookup/doi/10.1111/j.1365-2966.2011.19796.x.

Gunawardhana, M. L. P., A. M. Hopkins, J. Bland-Hawthorn, S. Brough, R. Sharp, J. Loveday, E. Taylor, et al. 2013. Galaxy And Mass Assembly: evolution of the Hα luminosity function and star formation rate density up to z < 0.35 [in en]. *Monthly Notices of the Royal Astronomical Society* 433, no. 4 (August): 2764–2789. ISSN: 1365-2966, 0035-8711, accessed March 14, 2023. https://doi.org/10.1093/mnras/stt890. http://academic.oup.com/mnras/article/433/4/2764/1747385/Galaxy-And-Mass-Assembly-evolution-of-the-HS%5Calpha$.

Gunawardhana, M. L. P., A. M. Hopkins, R. G. Sharp, S. Brough, E. Taylor, J. Bland-Hawthorn, C. Maraston, et al. 2011. Galaxy and Mass Assembly (GAMA): the star formation rate dependence of the stellar initial mass function: IMF-SFR relationship [in en]. *Monthly Notices of the Royal Astronomical Society* 415, no. 2 (August): 1647–1662. ISSN: 00358711, accessed March 14, 2023. https://doi.org/10.1111/j.1365-2966.2011.18800.x. https://academic.oup.com/mnras/article-lookup/doi/10.1111/j.1365-2966.2011.18800.x.

Hopkins, A. M., A. J. Connolly, D. B. Haarsma, and L. E. Cram. 2001. Toward a Resolution of the Discrepancy between Different Estimators of Star Formation Rate [in en]. *The Astronomical Journal* 122, no. 1 (July): 288–296. ISSN: 00046256, accessed October 16, 2023. https://doi.org/10.1086/321113. https://iopscience.iop.org/article/10.1086/321113.

Hopkins, A. M., C. J. Miller, R. C. Nichol, A. J. Connolly, M. Bernardi, P. L. Gomez, T. Goto, et al. 2003. Star Formation Rate Indicators in the Sloan Digital Sky Survey [in en]. *The Astrophysical Journal* 599, no. 2 (December): 971–991. ISSN: 0004-637X, 1538-4357, accessed March 29, 2023. https://doi.org/10.1086/379608. https://iopscience.iop.org/article/10.1086/379608.

Kauffmann, Guinevere, Timothy M. Heckman, Christy Tremonti, Jarle Brinchmann, Stéphane Charlot, Simon D. M. White, Susan E. Ridgway, et al. 2003. The host galaxies of active galactic nuclei [in en]. *Monthly Notices of the Royal Astronomical Society* 346, no. 4 (December): 1055–1077. ISSN: 00358711, 13652966, accessed August 7, 2023. https://doi.org/10.1111/j.1365-2966.2003.07154.x. https://academic.oup.com/mnras/article-lookup/doi/10.1111/j.1365-2966.2003.07154.x.

Kennicutt, Robert C. 1998. STAR FORMATION IN GALAXIES ALONG THE HUBBLE SEQUENCE [in en]. *Annual Review of Astronomy and Astrophysics* 36, no. 1 (September): 189–231. ISSN: 0066-4146, 1545-4282, accessed February 26, 2024. https://doi.org/10.1146/annurev.astro.36.1.189. https://www.annualreviews.org/doi/10.1146/annurev.astro.36.1.189.

Kong, X., S. Charlot, J. Brinchmann, and S. M. Fall. 2004. Star formation history and dust content of galaxies drawn from ultraviolet surveys [in en]. *Monthly Notices of the Royal Astronomical Society* 349, no. 3 (April): 769–778. ISSN: 00358711, 13652966, accessed March 29, 2023. https://doi.org/10.1111/j.1365-2966.2004.07556.x. https://academic.oup.com/mnras/article-lookup/doi/10.1111/j.1365-2966.2004.07556.x.

Koyama, Yusei, Tadayuki Kodama, Masao Hayashi, Rhythm Shimakawa, Issei Yamamura, Fumi Egusa, Nagisa Oi, et al. 2015. Predicting dust extinction properties of star-forming galaxies from Hα/UV ratio [in en]. *Monthly Notices of the Royal Astronomical Society* 453, no. 1 (October): 879–892. ISSN: 0035-8711, 1365-2966, accessed October 3, 2023. https://doi.org/10.1093/mnras/stv1599. https://academic.oup.com/mnras/article-lookup/doi/10.1093/mnras/stv1599.

Lin, Yen-Hsing, Hiroyuki Hirashita, Peter Camps, and Maarten Baes. 2021. Geometry effects on dust attenuation curves with different grain sources at high redshift. *Monthly Notices of the Royal Astronomical Society* 507, no. 2 (October): 2755–2765. https://doi.org/10.1093/mnras/stab2242. arXiv: 2109.03072 [astro-ph.GA].

Liske, J., I. K. Baldry, S. P. Driver, R. J. Tuffs, M. Alpaslan, E. Andrae, S. Brough, et al. 2015. Galaxy And Mass Assembly (GAMA): end of survey report and data release 2 [in en]. *Monthly Notices of the Royal Astronomical Society* 452, no. 2 (September): 2087–2126. ISSN: 0035-8711, 1365-2966, accessed March 14, 2023. https://doi.org/10.1093/mnras/stv1436. https://academic.oup.com/mnras/article-lookup/doi/10.1093/mnras/stv1436.

Lu, Jiafeng, Shiyin Shen, Fang-Ting Yuan, Zhengyi Shao, Jinliang Hou, and Xianzhong Zheng. 2022. The Double Chip Cookie Model: Dust Geometry of Milky Way-like Disk Galaxies. *The Astrophysical Journal* 938, no. 2 (October): 139. https://doi.org/10.3847/1538-4357/ac92e9. arXiv: 2209.08515 [astro-ph.GA].

Meurer, Gerhardt R., Timothy M. Heckman, and Daniela Calzetti. 1999. Dust Absorption and the Ultraviolet Luminosity Density at z \raisebox-0.5ex\textasciitilde 3 as Calibrated by Local Starburst Galaxies. _Eprint: astro-ph/9903054, *The Astrophysical Journal* 521, no. 1 (August): 64–80. https://doi.org/10.1086/307523.

Narayanan, Desika, Charlie Conroy, Romeel Davé, Benjamin D. Johnson, and Gergö Popping. 2018. A Theory for the Variation of Dust Attenuation Laws in Galaxies. *The Astrophysical Journal* 869, no. 1 (December): 70. https://doi.org/10.3847/1538-4357/aaed25. arXiv: 1805.06905 [astro-ph.GA].

Natale, Giovanni, Cristina C. Popescu, Richard. J. Tuffs, Victor P. Debattista, Jörg Fischera, and Meiert W. Grootes. 2015. Predicting the stellar and non-equilibrium dust emission spectra of high-resolution simulated galaxies with DART-RAY. *Monthly Notices of the Royal Astronomical Society* 449, no. 1 (May): 243–267. https://doi.org/10.1093/mnras/stv286. arXiv: 1502.03315 [astro-ph.GA].

Orellana, G., N. M. Nagar, D. Elbaz, P. Calderón-Castillo, R. Leiton, E. Ibar, B. Magnelli, et al. 2017. Molecular gas, dust, and star formation in galaxies: I. Dust properties and scalings in ~1600 nearby galaxies [in en]. *Astronomy & Astrophysics* 602 (June): A68. ISSN: 0004-6361, 1432-0746, accessed April 2, 2025. https://doi.org/10.1051/0004-6361/201629009. http://www.aanda.org/10.1051/0004-6361/201629009.

Osterbrock, Donald E. 1989. *Astrophysics of Gaseous Nebulae and Active Galactic Nuclei.* University Science Books.

Pierini, D., K. D. Gordon, A. N. Witt, and G. J. Madsen. 2004. Dust Attenuation in Late-Type Galaxies. I. Effects on Bulge and Disk Components [in en]. *The Astrophysical Journal* 617, no. 2 (December): 1022–1046. ISSN: 0004-637X, 1538-4357, accessed March 29, 2023. https://doi.org/10.1086/425651. https://iopscience.iop.org/article/10.1086/425651.

Popescu, C. C., R. J. Tuffs, M. A. Dopita, J. Fischera, N. D. Kylafis, and B. F. Madore. 2011. Modelling the spectral energy distribution of galaxies. V. The dust and PAH emission SEDs of disk galaxies. *Astronomy and Astrophysics* 527 (March): A109. https://doi.org/10.1051/0004-6361/201015217. arXiv: 1011.2942 [astro-ph.CO].

Popesso, P, A Concas, L Morselli, G Rodighiero, A Enia, and S Quai. 2020. The dust and cold gas content of local star-forming galaxies [in en]. *Monthly Notices of the Royal Astronomical Society* 496, no. 3 (August): 2531–2541. ISSN: 0035-8711, 1365-2966, accessed October 3, 2023. https://doi.org/10.1093/mnras/staa1737. https://academic.oup.com/mnras/article/496/3/2531/5859504.

Qin, Jianbo, Xian Zhong Zheng, Stijn Wuyts, Zongfei Lyu, Man Qiao, Jia-Sheng Huang, Feng Shan Liu, et al. 2024. Understanding the universal dust attenuation scaling relation of star-forming galaxies [in en]. *Monthly Notices of the Royal Astronomical Society* 528, no. 1 (January): 658–675. ISSN: 0035-8711, 1365-2966, accessed February 23, 2024. https://doi.org/10.1093/mnras/stad3999. https://academic.oup.com/mnras/article/528/1/658/7503925.

Sachdeva, Sonali, and Biman B. Nath. 2022. Star-dust geometry main determinant of dust attenuation in galaxies. *Monthly Notices of the Royal Astronomical Society* 513, no. 1 (June): L63–L67. https://doi.org/10.1093/mnrasl/slac037. arXiv: 2204.03478 [astro-ph.GA].

Taylor, Edward N., Andrew M. Hopkins, Ivan K. Baldry, Michael J. I. Brown, Simon P. Driver, Lee S. Kelvin, David T. Hill, Aaron S. G. Robotham, Joss Bland-Hawthorn, D. H. Jones, R. G. Sharp, Daniel Thomas, Jochen Liske, Jon Loveday, Peder Norberg, J. A. Peacock, Steven P. Bamford, Sarah Brough, Matthew Colless, Ewan Cameron, Christopher J. Conselice, Scott M. Croom, C. S. Frenk, Madusha Gunawardhana, Konrad Kuijken, R. C. Nichol, H. R. Parkinson, S. Phillipps, K. A. Pimbblet, C. C. Popescu, Matthew Prescott, W. J. Sutherland, R. J. Tuffs, Eelco van Kampen, et al. 2011a. Galaxy And Mass Assembly (GAMA): stellar mass estimates: GAMA: stellar mass estimates [in en]. *Monthly Notices of the Royal Astronomical Society* 418, no. 3 (December): 1587–1620. ISSN: 00358711, accessed May 22, 2023. https://doi.org/10.1111/j.1365-2966.2011.19536.x. https://academic.oup.com/mnras/article-lookup/doi/10.1111/j.1365-2966.2011.19536.x.

Taylor, Edward N., Andrew M. Hopkins, Ivan K. Baldry, Michael J. I. Brown, Simon P. Driver, Lee S. Kelvin, David T. Hill, Aaron S. G. Robotham, Joss Bland-Hawthorn, D. H. Jones, R. G. Sharp, Daniel Thomas, Jochen Liske, Jon Loveday, Peder Norberg, J. A. Peacock, Steven P. Bamford, Sarah Brough, Matthew Colless, Ewan Cameron, Christopher J. Conselice, Scott M. Croom, C. S. Frenk, Madusha Gunawardhana, Konrad Kuijken, R. C. Nichol, H. R. Parkinson, S. Phillipps, K. A. Pimbblet, C. C. Popescu, Matthew Prescott, W. J. Sutherland, R. J. Tuffs, Eelco van Kampen, et al. 2011b. Galaxy And Mass Assembly (GAMA): stellar mass estimates. *Monthly Notices of the Royal Astronomical Society* 418, no. 3 (December): 1587–1620. https://doi.org/10.1111/j.1365-2966.2011.19536.x. arXiv: 1108.0635 [astro-ph.CO].

Tuffs, R. J., C. C. Popescu, H. J. Völk, N. D. Kylafis, and M. A. Dopita. 2004. Modelling the spectral energy distribution of galaxies. III. Attenuation of stellar light in spiral galaxies. *Astronomy and Astrophysics* 419 (June): 821–835. https://doi.org/10.1051/0004-6361:20035689. arXiv: astro-ph/0401630 [astro-ph].

Wang, L., P. Norberg, M. L. P. Gunawardhana, S. Heinis, I. K. Baldry, J. Bland-Hawthorn, N. Bourne, et al. 2016. GAMA/H-ATLAS: common star formation rate indicators and their dependence on galaxy physical parameters [in en]. *Monthly Notices of the Royal Astronomical Society* 461, no. 2 (September): 1898–1916. ISSN: 0035-8711, 1365-2966, accessed March 14, 2023. https://doi.org/10.1093/mnras/stw1450. https://academic.oup.com/mnras/article-lookup/doi/10.1093/mnras/stw1450.

Wijesinghe, D. B., E. da Cunha, A. M. Hopkins, L. Dunne, R. Sharp, M. Gunawardhana, S. Brough, et al. 2011. GAMA/H-ATLAS: the ultra-violet spectral slope and obscuration in galaxies: UV spectral slope and obscuration [in en]. *Monthly Notices of the Royal Astronomical Society* 415, no. 2 (August): 1002–1012. ISSN: 00358711, accessed March 14, 2023. https://doi.org/10.1111/j.1365-2966.2011.18615.x. https://academic.oup.com/mnras/article-lookup/doi/10.1111/j.1365-2966.2011.18615.x.

Witt, Adolf N., Harley A. Thronson Jr., and John M. Capuano Jr. 1992. Dust and the transfer of stellar radiation within galaxies [in en]. *The Astrophysical Journal* 393 (July): 611. ISSN: 0004-637X, 1538-4357, accessed August 29, 2024. https://doi.org/10.1086/171530. http://adsabs.harvard.edu/doi/10.1086/171530.